\documentclass[letter,journal]{IEEEtran}
\usepackage[table,xcdraw]{xcolor}
\usepackage{color,soul}
\usepackage{cite}

\ifCLASSINFOpdf
   \usepackage[pdftex]{graphicx}
  \graphicspath{{./Figures/}}
  \DeclareGraphicsExtensions{.pdf,.jpeg,.png}
\else
  \usepackage[dvips]{graphicx}
  \usepackage{amsmath}
  \graphicspath{{./Figures/}}
  \DeclareGraphicsExtensions{.eps}
\fi

\usepackage{tikz}
\usetikzlibrary{arrows,shapes,chains,calc,patterns,decorations.pathreplacing}
\usepackage{comment}
\usepackage{amsthm,amsmath,amsfonts,dsfont,bm}

\usepackage{enumitem}
\usepackage[linesnumbered,lined,ruled,norelsize]{algorithm2e}
\usepackage{algorithmic}
\SetAlFnt{\small}

\usepackage{array}
\usepackage{multirow}
\newcommand{\tabincell}[2]{\begin{tabular}{@{}#1@{}}#2\end{tabular}}  

\ifCLASSOPTIONcompsoc
  \usepackage[caption=false,font=normalsize,labelfont=sf,textfont=sf]{subfig}
\else
  \usepackage[caption=false,font=footnotesize]{subfig}
\fi

\usepackage{dblfloatfix}

\ifCLASSOPTIONcaptionsoff
  \usepackage[nomarkers]{endfloat}
 \let\MYoriglatexcaption\caption
 \renewcommand{\caption}[2][\relax]{\MYoriglatexcaption[#2]{#2}}
\fi

\usepackage{url}
\usepackage[framemethod=tikz]{mdframed}
\usepackage{hhline}
\usepackage{fancyhdr}
\usepackage{booktabs}
\usepackage{longtable}
\usepackage{mfirstuc}
\usepackage{amssymb}

\newcommand{\RR}{\mathbb{R}}
\newcommand{\dd}{\mathrm{d}}

\newcommand{\NN}{\mathbb{N}}

\newcommand{\K}{\mathrm{K}}

\newcommand{\F}{\mathcal{F}}  
\newcommand{\Lrm}{\mathrm{L}}   
\newcommand{\srm}{\mathrm{s}}   

\newcommand{\red}[1]{\textcolor{black}{#1}}

\definecolor{lightblue}{rgb}{0.9, 1, 1}
\definecolor{lightyellow}{rgb}{1, 1, 0.9}
\definecolor{lightpink}{rgb}{1, 0.9, 1}
\definecolor{lightgreen}{rgb}{0.9, 1, 0.9}
\definecolor{lightgray}{rgb}{0.66, 0.66, 0.66}
\newcommand{\lightred}[1]{\textcolor[rgb]{0.8745, 0.498, 0.498}{#1}} 

\makeatletter
\newcommand{\removelatexerror}{\let\@latex@error\@gobble}
\makeatother

\begin{document}
%
\title{An Efficient Safety-oriented Car-following Model for Connected Automated Vehicles Considering Discrete Signals}

\author{DianChao~Lin
	and~Li~Li$^{\dagger}$
	\thanks{DianChao Lin is with the School of Economics and Management, Fuzhou University, Fuzhou 350108, China (email: lindianchao@fzu.edu.cn)}
	\thanks{$^{\dagger}$ Corresponding author. Li Li is with the School of Civil Engineering, Fuzhou University, Fuzhou 350108, China (email: lili@fzu.edu.cn)}
}

\maketitle


\begin{abstract}
With the rapid development of Connected and Automated Vehicle (CAV) technology, limited self-driving vehicles have been commercially available in certain leading intelligent transportation system countries. When formulating the car-following model for CAVs, safety is usually the basic constraint. Safety-oriented car-following models seek to specify a safe following distance that can guarantee safety if the preceding vehicle were to brake hard suddenly. The discrete signals of CAVs bring a series of phenomena, including discrete decision-making, phase difference, and discretely distributed communication delay. The influences of these phenomena on the car-following safety of CAVs are rarely considered in the literature. This paper proposes an efficient safety-oriented car-following model for CAVs considering the impact of discrete signals. The safety constraints during both normal driving and a sudden hard brake are incorporated into one integrated model to eliminate possible collisions during the whole driving process. The mechanical delay information of the preceding vehicle is used to improve car-following efficiency. Four modules are designed to enhance driving comfort and string stability in case of heavy packet losses. Simulations of a platoon with diversified vehicle types demonstrate the safety, efficiency, and string stability of the proposed model. Tests with different packet loss rates imply that the model could guarantee safety and driving comfort in even poor communication environments.
\end{abstract}


\IEEEpeerreviewmaketitle

\section{Introduction} 
\label{S:Intro}
\IEEEPARstart{C}{ar-following}, which is one of the most basic microscopic traffic behaviors, has been studied for around seventy years \cite{pipes1953operational}. Traditional car-following studies focus on reproducing the following dynamics of human drivers in simulations \cite{herman1959traffic, kometani1959dynamic, hanken1967model, evans1977perceptual, gipps1981behavioural, kikuchi1992car}. Automated vehicle (AV) technologies developed in the last decades, however, make it possible to implement car-following models in the real world \cite{lim2021state}. The car-following model of an AV can output a deterministic acceleration for the vehicle to execute \cite{talebpour2016influence}, without the participation of human drivers. Commercial applications such as vehicles with functions of Adaptive Cruise Control (ACC) and Cooperative ACC (CACC), have gained lots of research attentions in recent years \cite{moon2009design, milanes2013cooperative, milanes2014modeling, horiguchi2014study, gong2019cooperative, makridis2021openacc, shi2021empirical}.

It is widely believed that AVs can significantly improve car-following safety and efficiency. On one hand, AVs can reduce human errors, which account for 90\% of crashes \cite{haghi2014assessment, karbasi2022investigating}. On the other hand, AVs can react much faster in information collection \& processing and decision making than human drivers. They are hence able to handle a closer following headway and improve efficiency. Moreover, compared with disconnected AVs (e.g., vehicles equipped with ACC) that only use sensors to collect information, the Connected AVs (CAVs) can obtain more kinds of data through wireless communication. Examples include the electronic throttle opening angle information \cite{li2016car,sun2019car,chen2021car}, vehicle intention\cite{yang2020automatic}, and the status of multiple downstream vehicles in front of the Preceding Vehicle (PV)\cite{ma2022multiple}. These data cannot be detected/measured by sensors directly, but only be obtained through communication. They can help further shorten the reaction time of the Following Vehicle (FV) through anticipation of the PV's behavior.

The AV car-following models can be classified into different categories according to their main optimization goals, such as Safety-Oriented Car-Following (SOCF) model, efficiency-oriented model, and string-stability-oriented model. Among all goals, safety is the most basic one and hence is incorporated in all models. Compared with SOCF models, other models consider safety in rather simple ways, such as setting a fixed minimum space or time headway \cite{li2006cooperative, sawant2010longitudinal}, \red{or calculating a simple safety distance based on average speed \cite{xiong2020communication}}. The problem with such a fixed value is that efficiency will be wasted if the value is too large, while safety will be violated if the value is too small. There are also some studies trying to optimize the safety-efficiency balancing of AVs \cite{liu2020optimizing}, their key philosophy is that safety can be sacrificed to a certain extent in order to improve efficiency. 

The main goal of the SOCF model is to eliminate potential collision hazards during car following. It seeks to specify a safe following distance for the FV if the PV were to brake hard "unpredictably" \cite{brackstone1999car}. 
	Although all SOCF models for CAVs assume a fixed communication interval, none have systematically considered the effects brought by discrete signals, such as discretely distributed communication delay. For CAVs, signals are sent and received in discrete time, and decisions are also made discretely. That is, even the time that signal takes to transmit from PV to FV can be continuously distributed, the time from the PV's signal sending moment to the FV's signal using moments is discretely distributed. However, to the best of the authors' knowledge, this fact has been ignored by all existing literature, and such neglect can lead to a collision during driving (an example will be shown in Fig. \ref{F:3_1}).

In this paper, we propose an efficient SOCF model for CAVs with discrete signals. Our principal contributions are summarized in the following.
\begin{enumerate}[label=\arabic*)]
	\item We systematically consider the influence of discrete signals on CAV car-following to reproduce the real communication and decision-making process. Specifically, we consider the phase difference, and the discretely distributed communication delay resulting from random transmission delay, packet loss, and discrete decision-making.
	\item We integrate the real-time space headway constraint and the safe distance constraints in case of an emergency hard brake into one acceleration decision-making model. This guarantees car-following safety during the whole driving process of CAVs with discrete signals.
	\item We use the mechanical delay of PV as one kind of communicated information to predict PV's status within a short period. This helps reduce the following headway and improve efficiency. 
	\item We design four modules in the simulation to improve driving comfort and string stability of platoons in case of poor communication environments with high packet loss rates. Their effectiveness is validated through simulations.
\end{enumerate}

The rest of this paper is organized as follows: the next section reviews the related literature about SOCF; the third section formulates the methodology; the fourth section presents the simulation results; and the last section concludes the paper.

\section{Literature Review} 
\label{S:litr}
The SOCF model originated from Kometani's work \cite{kometani1959dynamic}, and was then majorly developed by Gipps \cite{gipps1981behavioural}. It seeks to guarantee driving safety during the whole car-following process, even when PV brakes hard "unpredictably". A SOCF model usually constructs its safety constraints by considering that in case of a sudden hard brake of PV, the safest maneuver FV can take is to brake hard immediately after a delay. For a human driver, the delay includes visual response time, decision-making time, decision execution time (e.g., accelerator release, foot movement from the accelerator to the brake pedal, and pedal pushing), and mechanical delay. The overall delay can range from 0.75 s to 4 s \cite{gipps1981behavioural,macadam2003understanding}. For a CAV, the delay includes information transmission and queueing time, decision-making time, and mechanical delay. The overall delay for a CAV can be as short as 0.2 s \cite{rajamani2011vehicle}.


The safety constraints of SOCF model have experienced several developments. The early-version SOCF model for human drivers only constrains the two vehicles not to collide when they both come to a full stop after brake\cite{gipps1981behavioural}. However, evidence in some safety evaluations \cite{touran1999collision} research has shown that when the braking performance of FV is better than PV, collisions could happen in the midway of the braking process even when safety at a full stop is guaranteed. Later, a SOCF model developed for disconnected (dis-C) AVs called the Responsibility-Sensitive Safety (RSS) model \cite{shalev2017formal} is proposed. RSS model still only constrains the safety at a full stop but assumes that FV always brakes more slowly (with a smaller absolute value of deceleration rate) than PV. Such assumption indeed eliminates the collision risk in a continuous decision-making system, however, it significantly sacrifices the car-following efficiency. This is because vehicles have to keep a much larger headway than needed, especially when a light vehicle follows a heavy vehicle. Recently, SOCF models for CAVs relax this assumption and consider the midway collision risk during brake when formulating the safety constraints \cite{dollar2018efficient,zhao2019dsrc,li2020car,hu2021modeling}. Compared with the RSS model, this modification significantly improves efficiency. Compared with Gipps Model, this modification improves the safety in some circumstances by eliminating the midway collisions during brake.

Discrete and periodical information transmission and control decision yield a series of phenomena, such as the phase difference (the time difference between two adjustment moments of PV and FV) and the discretely distributed communication delay (information's transmission and queueing time from PV to FV). Furthermore,  
packet loss increases the uncertainty of information transmission and the risk (and discomfort) of a CAV's traveling. These challenge vehicle platoon's safety and string stability of the state-of-art CAV SOCF models \cite{dollar2018efficient,zhao2019dsrc,li2020car,hu2021modeling}.

\begin{table}[h!]
	\caption{Comparison of main SOCF research and this paper}
	\centering
	\setlength\tabcolsep{1.75pt}
	\setlength{\extrarowheight}{3pt}
	\label{T:lit}
		\begin{tabular}{c|cccccccc}
			\hline
			Paper                                       & \begin{tabular}[c]{@{}c@{}}Traffic\\ envr.\end{tabular} & \begin{tabular}[c]{@{}c@{}}Have\\ comm.\\ cycle?\end{tabular} & \begin{tabular}[c]{@{}c@{}}Have\\ control\\ cycle?\end{tabular} & \begin{tabular}[c]{@{}c@{}}Random\\ (comm.)\\ delay?\end{tabular} & \begin{tabular}[c]{@{}c@{}}Send\\ mech.\\ info.?\end{tabular} & \begin{tabular}[c]{@{}c@{}}Packet\\ loss?\end{tabular} & \begin{tabular}[c]{@{}c@{}}Veh.\\ type\end{tabular} & \begin{tabular}[c]{@{}c@{}}Safety\\ cstr.\end{tabular}                                     \\ \hline
			\tabincell{c}{Gipps \\ \cite{gipps1981behavioural}}                        & HDV                                                     & No                                                           & No                                                              & No                                                             & No                                                          & No                                                     & 1                                                      & full stop                                                                                 \\ \hline
			\tabincell{c}{Moon\\\cite{moon2009design}}                              & \tabincell{c}{dis-C\\AV+\\HDV}                                                  & No                                                           & Yes                                                             & No                                                             & No                                                          & No                                                     & 1                                                      & full stop                                                                                 \\ \hline
			\tabincell{c}{Milanes\\\cite{milanes2013cooperative, milanes2014modeling}} & CAV                                                     & Yes                                                          & Yes                                                             & No                                                             & No                                                          & No                                                     & 1                                                      & \begin{tabular}[c]{@{}c@{}}constant\\ time \\ headway\end{tabular}                           \\ \hline
			Guo\cite{guo2021anticipative}                         & \tabincell{c}{CAV+\\HDV}                                                 & Yes                                                          & Yes                                                             & \begin{tabular}[c]{@{}c@{}}Yes, \\ cont.\end{tabular}     & No                                                          & 10-40\%                                                & 1                                                      & \begin{tabular}[c]{@{}c@{}}constant\\ space \\ headway\end{tabular}                          \\ \hline
			\tabincell{c}{Shalev-\\
				Shwartz\\ \cite{shalev2017formal},\\
				Liu\cite{liu2021calibration}}        & \tabincell{c}{dis-C\\AV}                                                & No                                                           & No                                                              & No                                                             & No                                                          & No                                                     & $>1$                                        & full stop                                                                                 \\ \hline
			\tabincell{c}{Dollar\\ \cite{dollar2018efficient}}                         &  \tabincell{c}{CAV+\\HDV}                                                 & Yes                                                          & Yes                                                             & No                                                             & No                                                          & 1-20\%                                                 & \tabincell{c}{small\\large}                                                      & \begin{tabular}[c]{@{}c@{}}midway \\ brake \&\\ full stop\end{tabular}                       \\ \hline
			\tabincell{c}{Zhao\cite{zhao2019dsrc},\\Li\cite{li2020car},\\Hu\cite{hu2021modeling}}      & CV                                                      & Yes                                                          & No                                                              & \begin{tabular}[c]{@{}c@{}}Yes,\\ cont.\end{tabular}      & No                                                          & No                                                     & 1                                                      & \begin{tabular}[c]{@{}c@{}}midway \\ brake \&\\ full stop\end{tabular}                       \\ \hline
			\tabincell{c}{This\\paper}                                  & CAV                                                     & Yes                                                          & Yes                                                             & \begin{tabular}[c]{@{}c@{}}Yes,\\ discrete\end{tabular}        & Yes                                                         & 0-50\%                                                 & 3                                                      & \begin{tabular}[c]{@{}c@{}}real-time\\ +midway \\ brake \& \\ full stop\end{tabular} \\ \hline
		\end{tabular}
\end{table} 

Table \ref{T:lit} lists the comparison between the main literature in SOCF and this paper. As we can see, existing research either considered fixed communication delay or continuously distributed random delay, this paper formulates the model based on the discretely distributed delay for the first time (the first contribution). Besides, current studies either consider the real-time headway constraint or safety constraint(s) in case of a hard brake, while this paper integrates both (second contribution). In addition, PV's mechanical delay is sent and used for the first time in this paper, which helps FV better deduce PV's current status (third contribution). Finally, although some studies have considered packet loss, they only proposed simplified interpolation methods to make predictions about PV, this paper designs four detailed modules to improve driving comfort and string stability in case of packet loss (fourth contribution).

\section{Methodology}
To avoid obscurity between a superscript and a square in this paper, we use a bracket to separate a parameter and its square, like $(\cdot)^2$. Besides, we use an overline to represent a variable's upper bound and an underline to represent its lower bound.

\subsection{Safety Constraint}
\label{ss:sm}
We number the CAVs in a platoon in ascending order from the first to the last. Let $l_n$ denote the length of CAV $n$, and $s_n$ denote the minimum gap that $n$ wants to keep away from its PV $n-1$' tail. 
For any time $t$, $x_n^t$, $\dot{x}_n^t$, and $\ddot{x}_n^t$ represent the head position, speed, and acceleration of CAV $n$ respectively. $y_n^{t_1 \rightarrow t}$ denotes the moving distance of CAV $n$ from time $t_1$ to $t$ if it were to brake hard during this time period. 

\begin{figure}[h!] \vspace{-1mm}
	\centering
	\includegraphics[width=0.95\linewidth]{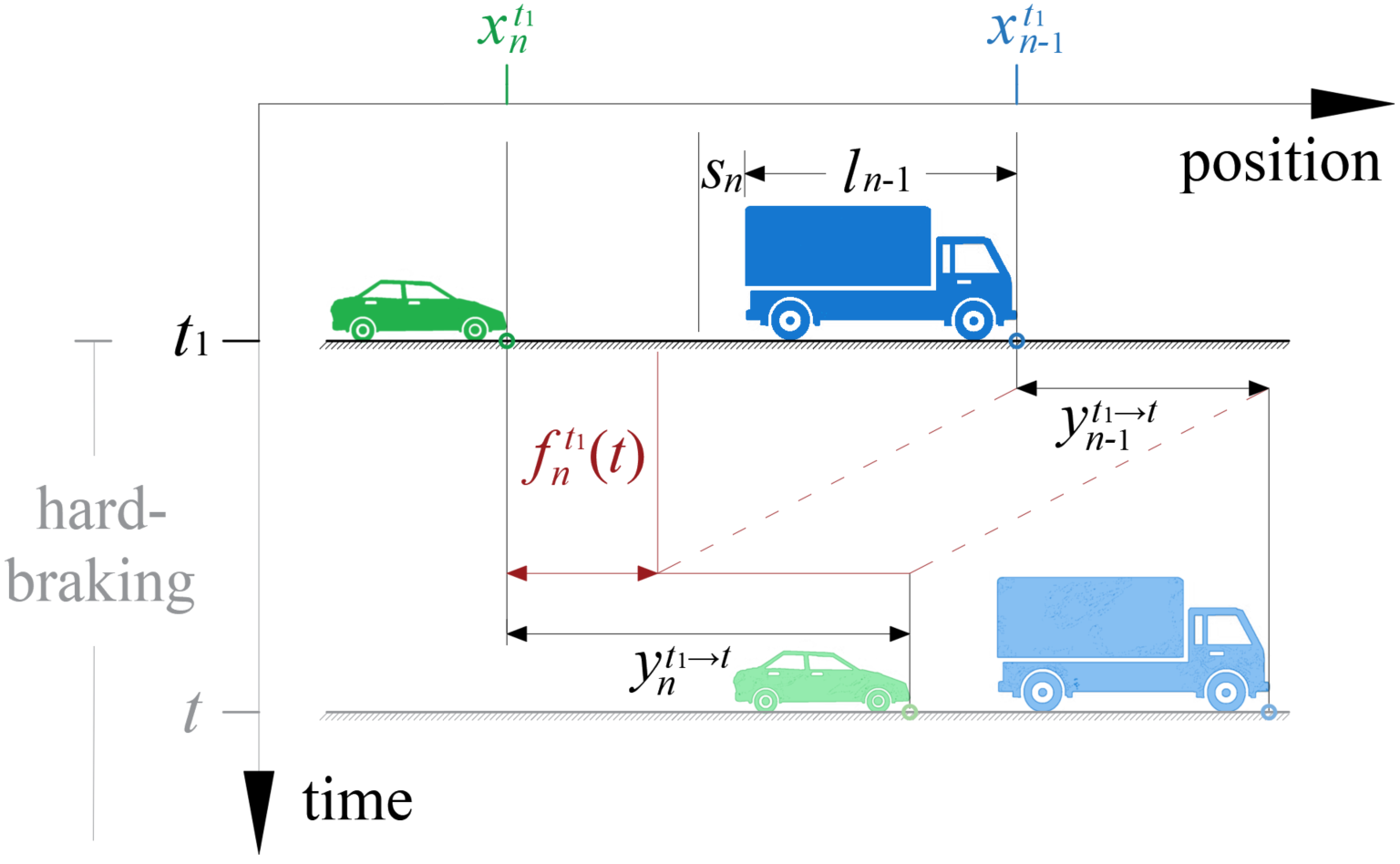}
	\caption{Safety consideration for vehicle $n$ behind $n-1$ at time $t_1$.}
	\label{fig:safe_margin}
\end{figure}

Fig. \ref{fig:safe_margin} shows a scenario in which 
a car $n$, behind a truck $n-1$, is determining its following behavior at time $t_1$. 
If $n-1$ brakes hard from $t_1$, the safest action that $n$ can take is to also brake hard from $t_1$. We show this imagined hard-braking by lighter color in Fig. \ref{fig:safe_margin}. Clearly, at any time $t\geq t_1$, the distance two vehicles approach, denoted by $f_n^{t_1}(t)$, is
\begin{equation}
	f_n^{t_1}(t) = y_n^{t_1 \rightarrow t} - y_{n-1}^{t_1 \rightarrow t}.
	\label{E:f}
\end{equation}	
To guarantee safety, we have the safety constraint
	\begin{equation}
	f_n^{t_1}(t) + s_n + l_{n-1} \leq x_{n-1}^{t_1} - x_{n}^{t_1}, ~~ \forall t \in [t_1, t_\mathrm{s}],
	\label{NE:safe_constraint}
\end{equation}
where $t_\mathrm{s}$ is the moment when both hard-braking vehicles have just stopped. \eqref{NE:safe_constraint} prevents rear-end collisions from the start-point of a brake to a full stop.

It is worth mentioning that, we only consider two vehicles when formulating the safety constraints in this paper. Although considering more vehicles (e.g., a head vehicle of a platoon) may reduce FV's reaction time, using information from more than PV also brings additional risks. For example, when facing a hacker attack \cite{liu2022distributed}, serial rear-end collisions may easily happen if the head vehicle is hacked since its information is used by many FVs in the platoon. However, if each FV only uses its direct PV’s information, the hacker needs much more effort to cause serial rear-end collisions. Hence, considering only two vehicles is also for cyber security reasons.

\subsection{Delays}
\begin{figure}[h!]
	\centering
	\includegraphics[width=0.95\linewidth]{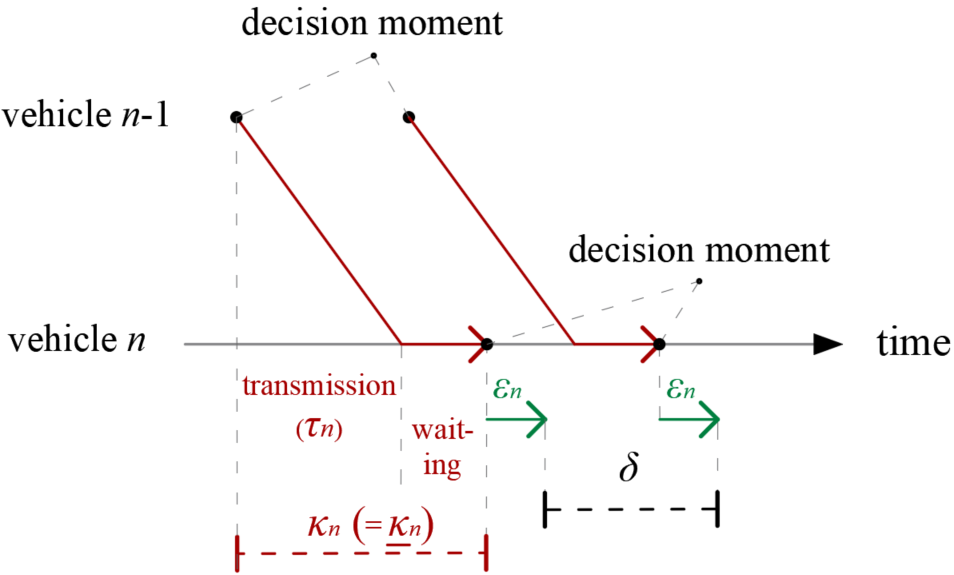}
	\caption{Communication and execution of CAV}
	\label{fig:timemap}
\end{figure}

In a connected vehicle environment, vehicles transmit their status information every communication cycle, denoted as $\delta$. In this paper, we make vehicles communicate their mechanical delays in addition to the basic information (e.g., acceleration, velocity, position, \textit{etc.}). Based on the received information from other vehicles, each vehicle makes the acceleration decision periodically and then broadcasts the latest decision (and other related information) along with the corresponding mechanical delay. For each vehicle, the mechanical delay can be measured offline from real-world experiments in the form of a look-up table for an online application \cite{wang2018delay} or be directly estimated online by a delay observer \cite{chang1995time}. Usually, the decision-making interval is the same length as the communication interval. That is, every $\delta$, vehicles receive information, make decisions, and then send information, as shown in Fig. \ref{fig:timemap}.
The information from $n-1$ to $n$ could experience multiple delays. Using a specific communication protocol, information is coded and compressed by $n-1$, then sent to and decompressed by $n$. To simplify the analysis, we merge these delays into a single transmission delay denoted by $\tau_n$. The decompressed information will wait for some time before being used by vehicle $n$ for decision making. We name the total delay caused by transmission and waiting as communication delay, which is denoted by $\kappa_n$. Note that $\underline{\kappa}_n$ is $\kappa_n$'s lower bound (with the smallest waiting time). Due to the randomness in transmission delay and packet loss, $\kappa_n$ is not necessarily equal to $\underline{\kappa}_n$ considering the driving comfort. The decision delay \red{(denoted by $d_n$)} after $\kappa_n$ is usually very short, hence without loss of generality, it is set to zero in this paper. The determined acceleration will be transmitted to $n$'s FV (if any), and meanwhile experience a mechanical delay, denoted by $\epsilon_n$, before being executed by $n$. Again, without loss of generality, we assume that every acceleration decision will be conducted for a time interval of $\delta$ length. That is, the acceleration during each $\delta$ is constant.

\begin{figure}[h!]
	\centering
	\includegraphics[width=0.9\linewidth]{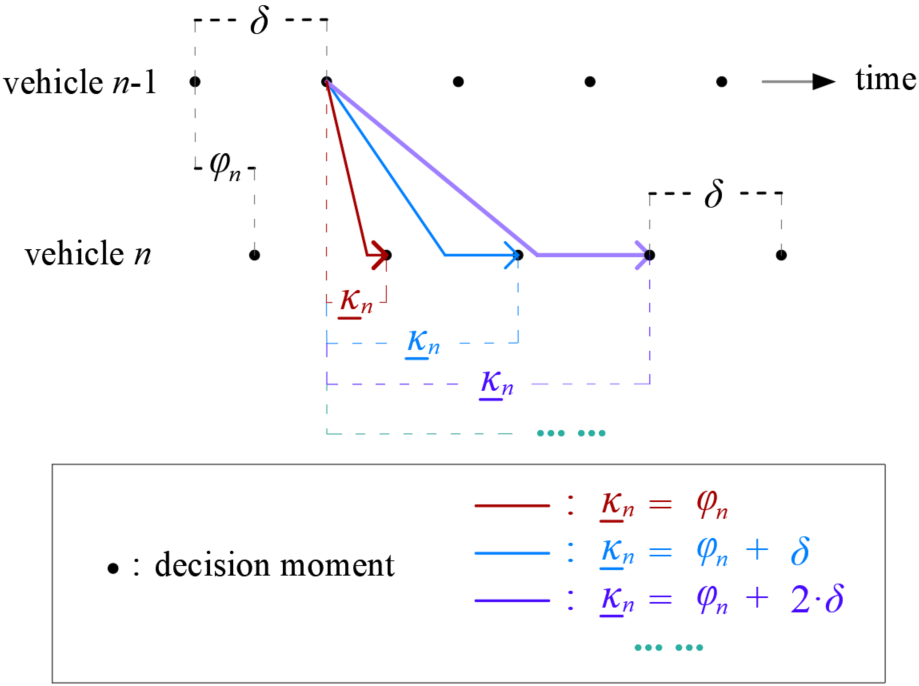}
	\caption{Discrete distribution of $\underline{\kappa}_n$.}
	\label{fig:kappa}
\end{figure}

For $n$'s arbitrary decision moment (denoted by $t_0$), its decision result will be executed during time interval ($t_1-\delta$, $t_1$], where $t_1 = t_0 + \epsilon_n + \delta$. If that $n$'s decision at $t_0$ uses $n-1$'s information decided and sent at $\widetilde{t}_0$, because of the existence of mechanical delay, the corresponding latest status of $n-1$ is at time $\widetilde{t}_1 = \widetilde{t}_0 + \epsilon_{n-1} + \delta$. From Fig. \ref{fig:timemap}, we have that $t_0 = \widetilde{t}_0 + \kappa_n$. Note that the lower bound of communication delay $\underline{\kappa}_n$ may fluctuate drastically due to the randomness in transmission delay $\tau_n$. And because of the discreteness of decision moments, $\underline{\kappa}_n$ follows a discrete distribution. 
Fig. \ref{fig:kappa} shows how $\tau_n$ affects $\underline{\kappa}_n$. $\phi_n$ in Fig. \ref{fig:kappa} is the phase difference of decision moments between $n$ and $n-1$. 
If $\phi_n + (\nu -1) \cdot \delta < \tau_n \leq \phi_n + \nu \cdot \delta$, where $\nu \in \NN$, we have $\underline{\kappa}_n = \phi_n + \nu \cdot \delta$. Once $\underline{\kappa}_n$ is determined, we can choose $\kappa_n = \underline{\kappa}_n + \mu_n \cdot \delta$ as needed, where $\mu_n \in \NN$. Note that we do not always choose $\kappa_n = \underline{\kappa}_n$ because sometimes a delayed use of information is needed to smooth the acceleration fluctuation, which will be detailed in the simulation section. \red{Note that, in some complicated traffic environments where the duration of decision delay $d_n$ cannot be ignored (due to limited computation resource), we have $\underline{\kappa}_n = \phi_n + \nu \cdot \delta$ when $\phi_n + (\nu -1) \cdot \delta < \tau_n + d_n \leq \phi_n + \nu \cdot \delta$.} 

In case of a packet loss, as shown in Table \ref{T:sample_pl}, the lost information from the PV $n-1$ (gray) is regarded with infinite transmission delay and hence will never reach the FV $n$. When making a decision, $n$ has to either use the old information or use neighboring information to estimate the lost information. Details will be found in the simulation section. 
\begin{table}[h!]  
	\caption{An example of information transmission and packet loss (gray) ($\phi_n=0.05 ~ \mathrm{s}$, $\tau_n \in [0.04,0.08] ~ \mathrm{s}$, and $\delta= 0.1 ~ \mathrm{s}$).}
	\label{T:sample_pl}
	\tabcolsep=0.08cm
	\begin{tabular}{ccccccc}
		\multicolumn{2}{c}{Info $n-1$ sends}               				&                      & \multicolumn{4}{c}{Info $n$ receives}                                                    \\ \hhline{*{2}{-}~|*{4}{-}} 
		
		\multicolumn{1}{|c}{\tabincell{c}{\textbf{~Sent~}\\ \textbf{~~time~} \\}}& \multicolumn{1}{c|}{\tabincell{c}{\textbf{~Status~}\\ \textbf{~~info~} \\}}& \multicolumn{1}{c|}{$\bm{\tau_n}$:} & \multicolumn{1}{c}{\tabincell{c}{\textbf{~Sent~}\\ \textbf{~~time~} \\}}& \tabincell{c}{\textbf{~Received~}\\ \textbf{~~time~}\\} & \multicolumn{1}{c}{\tabincell{c}{\textbf{Min propagation}\\ \textbf{delay} \\}}  & \multicolumn{1}{c|}{\tabincell{c}{\textbf{~Status~}\\ \textbf{~~info~} \\}}               \\ \hhline{*{2}{-}~|*{4}{-}} 
		
		\multicolumn{2}{|c|}{$\dots$}    & \multicolumn{1}{c|}{}                         &    \multicolumn{4}{c|}{$\dots$}                    \\ \hhline{*{2}{-}~|*{4}{-}}
		  
		\multicolumn{1}{|c}{20.0}                           & \multicolumn{1}{c|}{$\dots$}     & \multicolumn{1}{c|}{$\xrightarrow{0.069~}$}    &    20.0   & 20.069            &      0.15                   & \multicolumn{1}{c|}{$\dots$}                  \\ \hhline{*{2}{-}~|*{4}{-}} 
		
		\multicolumn{1}{|c}{20.1}                    &\multicolumn{1}{c|}{$\dots$}     & \multicolumn{1}{c|}{$\xrightarrow{0.045~}$}        &  \multicolumn{1}{c}{20.1}   & 20.145               &     0.05                        & \multicolumn{1}{c|}{$\dots$}                  \\ \hhline{*{2}{-}~|*{4}{-}} 
		
		\multicolumn{1}{|c}{\cellcolor{lightgray}20.2}                    &  \multicolumn{1}{c|}{\cellcolor{lightgray}$\dots$}      &   \multicolumn{1}{c|}{$\xrightarrow{~~\bm{\infty}~~}$}   & \cellcolor{lightgray}20.2  & \cellcolor{lightgray}?	& \cellcolor{lightgray}? &   \multicolumn{1}{c|}{\cellcolor{lightgray}?}                 \\ \hhline{*{2}{-}~|*{4}{-}} 
		
		\multicolumn{1}{|c}{20.3}                         &\multicolumn{1}{c|}{$\dots$}  & \multicolumn{1}{c|}{$\xrightarrow{0.053~}$} & \multicolumn{1}{c}{20.3}  &  20.353    &   0.15             &  \multicolumn{1}{c|}{$\dots$}  \\ \hhline{*{2}{-}~|*{4}{-}|} 
		
		\multicolumn{2}{|c|}{$\dots$}          & \multicolumn{1}{c|}{}                            &               \multicolumn{4}{c|}{$\dots$}               \\ \hhline{*{2}{-}~|*{4}{-}} 
	\end{tabular} \vspace{-3mm}
\end{table} 

\subsection{Conservative Safety Constraint} 

When $n$ decides its acceleration for time interval ($t_1-\delta$, $t_1$], it tends to use $n-1$'s status information at $t_1$ to guarantee the safety for this interval. However, because of the information delays, if $n-1$'s latest status information (that $n$ could use) at time $\widetilde{t}_1<t_1$, then $n$ has to estimate $n-1$'s status from $\widetilde{t}_1$ to $t_1$.  Conservatively, $n-1$ may brake hard from $\widetilde{t}_1$ to $t_1$, which is the scenario with the minimum moving distance during interval $(\widetilde{t}_1$, $t_1]$. Similar to the analysis in \ref{ss:sm}, we define a \emph{conservative difference of hard-braking distance,} $\F_n^{t_1}(t)$, as
	\begin{equation}
		\F_n^{t_1}(t) = y_n^{t_1 \rightarrow t} - \big(y_{n-1}^{t_\K \rightarrow t} - y_{n-1}^{t_\K \rightarrow t_1} \big) \ge f_n^{t_1}(t),
		\label{E:F}
	\end{equation}
where $t \ge t_1$, and $t_\K := \min\big\{t_1,\widetilde{t}_1\big\}$. When $t_1 \leq \widetilde{t}_1$,we have $t_\K = t_1$, and \eqref{E:F} becomes the same with \eqref{E:f} with a conservative assumption that vehicle $n-1$ starts to brake hard from $t_1$ (the knowledge of $n-1$'s status from $t_1$ to $\widetilde{t}_1$ is not used for simplicity). When $\widetilde{t}_1 < t_1$, we have $t_\K = \widetilde{t}_1$, and \eqref{E:F} assumes $n-1$ brakes hard from $\widetilde{t}_1$. Hence, the hard-braking distance of $n-1$ from $t_1$ to $t$ is equal to the hard-braking distance from $\widetilde{t}_1$ to $t$ minus the hard-braking distance from $\widetilde{t}_1$ to $t_1$. We note that without the knowledge of $n-1$'s mechanical delay ($\epsilon_{n-1}$), $n$ will need to make a conservative assumption that $n-1$ brakes hard from time $\widetilde{t}_1 - \epsilon_{n-1}$. Hence a larger following space will be imposed on $n$ than actually needed.

Moreover, \red{results in the literature suggest keeping acceleration fluctuations as small as possible to improve comfort in automated driving \cite{bellem2018comfort}. Therefore, to reduce} the acceleration fluctuations due to phase difference, random transmission delay, and packet loss \red{to enhance driving comfort}, we design an \emph{elastic minimum gap}, $S_n$, as
\begin{equation}
	S_n = \gamma \cdot \delta \cdot \dot{x}_n^{t_1} + s_n,	\label{E:overline_s}
\end{equation}
where $\gamma \geq 0$. Compared with $s_n$, $S_n$ includes an extra gap which increases linearly with $n$'s speed.

Therefore, similar to \eqref{NE:safe_constraint}, we have a conservative safety constraint given as
\begin{equation}
	 \F_n^{t_1}(t) + S_n + l_{n-1} \leq x_{n-1}^{t_\K} + y_{n-1}^{t_\K \rightarrow t_1} - x_n^{t_1}, \quad \forall t \in [t_1, t_\mathrm{s}],
	\label{Ine:dist_b}
\end{equation}
where again, $t_s$ is the stop moment for both vehicles in the corresponding hard-braking scenario. Similar to \eqref{NE:safe_constraint}, \eqref{Ine:dist_b} excludes any rear-end collisions from the start-point of any potential brake to full stop. Satisfying \eqref{Ine:dist_b} in real-time ensures safety during both normal driving and any sudden brake.

Because $\F_n^{t_1}(t)$ is continuous (and differentiable) on a closed interval, its maximum always exists. Hence, \eqref{Ine:dist_b} is true iff  $\underset{t \in [t_1, t_\srm]}{\max}\big\{\F_n^{t_1}(t)\big\} + S_n + l_{n-1} \leq x_{n-1}^{t_\K} + y_{n-1}^{t_\K \rightarrow t_1} - x_n^{t_1}$ is true. However, it is difficult to determine $\F_n^{t_1}(t)$'s maximum because its location changes with the decision variable $\ddot{x}_n^{t_1}$. As global maximum point could be at $t_1$, $t_\srm$, or $t_\Lrm$ (where $t_\Lrm \in (t_1, t_\srm)$ is the \emph{location of a local maximum point} excluding the start-point and end-point) if $t_\Lrm$ exists.  Therefore, \eqref{Ine:dist_b} is equivalent to the inequality group: 
\begin{equation}
\begin{cases} 
	\F_n^{t_1}(t_1) + S_n + l_{n-1} \leq x_{n-1}^{t_\K} + y_{n-1}^{t_\K \rightarrow t_1} - x_n^{t_1},  \\
	\F_n^{t_1}(t_\srm) + S_n + l_{n-1} \leq x_{n-1}^{t_\K} + y_{n-1}^{t_\K \rightarrow t_1} - x_n^{t_1},  \\
	\F_n^{t_1}(t_\Lrm) + S_n + l_{n-1} \leq x_{n-1}^{t_\K} + y_{n-1}^{t_\K \rightarrow t_1} - x_n^{t_1} ~~ \mathrm{if}~ t_\Lrm~\mathrm{exits.}\\
\end{cases}
\label{Ine:dist_c}
\end{equation}

We will discuss the three safety constraints in \eqref{Ine:dist_c} in detail later when calculating the domain of feasible accelerations.

\subsection{Update of Status}
Since $n$'s the acceleration (the decision variable) is constant for each interval $(t_1-\delta, t_1]$, we can update the position and speed of $n$ at $t_1$ as
\begin{multline}
	~~~~~~\begin{bmatrix} 
		x_n^{t_1} \\ 	\dot{x}_n^{t_1} 
	\end{bmatrix}
	=
	\begin{bmatrix} 
		1 & \delta & 0.5 (\delta)^2\\ 0 & 1 & \delta  
	\end{bmatrix} 
	\begin{bmatrix} 
		~~~x_n^{t_1-\delta} \\ ~~~\dot{x}_n^{t_1-\delta} \\ 
		\ddot{x}_n^{t_1}
	\end{bmatrix}. ~~~~~~~~~
	\label{E:B1}
\end{multline}
If $n$ were to brake hard from $t_1$ to $t$, its braking distance from $t_1$ to $t$ ($y_n^{t_1 \rightarrow t}$) and braking speed at $t$ (denoted by $\dot{y}_n^{t_1 \rightarrow t}$) can be calculated as
	\begin{multline}
		~~~~\begin{bmatrix} 
			y_n^{t_1 \rightarrow t} \\ 	\dot{y}_n^{t_1 \rightarrow t} 
		\end{bmatrix}
		=
		\begin{bmatrix} 
			1 & \Delta t & 0.5 (\Delta t)^2\\ 0 & 1 & \Delta t 
		\end{bmatrix} 
		\begin{bmatrix} 
			0 \\ \dot{x}_n^{t_1} \\ 
			\underline{\ddot{x}}_n
		\end{bmatrix}, ~~~~~
		\label{E:B2}
	\end{multline}
where $\underline{\ddot{x}}_n$ is the lower bound of $n$'s acceleration (negative), and $\Delta t = \min\big\{t - t_1, \dot{x}_n^{t_1}/-\underline{\ddot{x}}_n \big\}$. Status of $n-1$ can be updated similarly and hence will not be repeated here.

\subsection{Determination on Acceleration}
\label{SS:acc}
We now discuss how to determine $\ddot{x}_n^{t_1}$ at every decision moment. Before considering the safety car-following constraints in \eqref{Ine:dist_c}, we need to supplement some basic constraints.

\subsubsection{Acceleration domain for basic constraints}
First of all, $\ddot{x}_n^{t_1}$ should be constrained by the acceleration and speed boundaries:
\begin{align}
	\underline{\ddot{x}}_n  \leq & \ddot{x}_n^{t_1} \leq \overline{\ddot{x}}_n,
	\label{E:acceleration_4a} \\
	\mathrm{and} \quad \frac{-\dot{x}_n^{t_1-\delta}}{\delta}  \leq & \ddot{x}_n^{t_1} \leq \frac{\overline{\dot{x}}_n-\dot{x}_n^{t_1-\delta}}{\delta}, \quad
	\label{E:acceleration_4b}
\end{align}
where $\overline{\ddot{x}}_n$ is the upper bound of $n$'s acceleration, and $\overline{\dot{x}}_n$ is the upper bound of $n$'s speed. We denote the resulted basic domain by $\bm{\Lambda^0}$, with
\begin{equation}
	\bm{\Lambda^0} = \big\{ \ddot{x}_n^{t_1} \big|  \eqref{E:acceleration_4a}, \eqref{E:acceleration_4b} \big\}.
\end{equation}

\subsubsection{Acceleration domain at the start-point of a potential hard brake}
In the 1st constraint of \eqref{Ine:dist_c}, $t = t_1$, and $\F_n^{t_1}(t_1) = 0$. Combined with \eqref{E:overline_s}, \eqref{E:B1} and \eqref{E:B2}, we have
\begin{multline}
	\ddot{x}_n^{t_1} \leq \frac{2}{(2\gamma+1)(\delta)^2} \big( x_{n-1}^{t_\K} + \dot{x}_{n-1}^{t_K}\theta  + \frac{\underline{\ddot{x}}_{n-1}}{2}(\theta)^2 \\ - x_n^{t_1-\delta} - (\gamma+1)\dot{x}_n^{t_1-\delta}\delta - l_{n-1} - s_n \big), 
	\label{E:acceleration_1}
\end{multline}
where 
\begin{equation}
	\theta = \min\{t_1 - t_\K, \frac{\dot{x}_{n-1}^{t_\K}}{-\underline{\ddot{x}}_{n-1}} \}.
\end{equation}	
We denote the resulted feasible domain of \eqref{E:acceleration_1} by $\bm{\Lambda^{t_1}}$, with
\begin{equation}
	\bm{\Lambda^{t_1}} = \big\{ \ddot{x}_n^{t_1} \big| \eqref{E:acceleration_1} \big\}.
\end{equation}
We name the safety constraint $\bm{\Lambda^{t_1}}$ as \emph{start-point constraint} since it considers the safety at the start-point of a potential hard brake. Note that $\bm{\Lambda^{t_1}}$ constrains the space between PV and FV before a potential brake actually happens, hence it is effectively a real-time following space constraint that FV needs to obey. Compared with the fixed space constraint set in some efficiency-oriented car-following models \cite{li2006cooperative, sawant2010longitudinal} in the literature, $\bm{\Lambda^{t_1}}$ is dynamically adjusted according to the real-time status of PV and FV. Hence it can guarantee safety and improve efficiency as well.

\subsubsection{Acceleration domain at the end-point of a potential hard brake}
In the 2nd constraint of \eqref{Ine:dist_c}, $t = t_\srm$. Because the vehicle who stops earlier will keep a speed of 0 after its stop, we have $\mathcal{F}_n^{t_1}(t_\srm) = \frac{(\dot{x}_n^{t_1})^2}{-2\underline{\ddot{x}}_n}-\frac{(\dot{y}_{n-1}^{t_\K\rightarrow t_1})^2}{-2\underline{\ddot{x}}_{n-1}}$. Combined with \eqref{E:overline_s}, \eqref{E:B1} and \eqref{E:B2}, we have
\begin{equation}
	\frac{-\alpha_1 -\sqrt{(\alpha_1)^2 -4\alpha_2}}{2}  \leq \ddot{x}_n^{t_1} \leq \frac{-\alpha_1 +\sqrt{(\alpha_1)^2 -4\alpha_2}}{2},
	\label{E:acceleration_2}
\end{equation}
where 
\begin{equation}
	\alpha_1 = \frac{2 \dot{x}_n^{t_1-\delta}}{\delta} - (2\gamma+1)\underline{\ddot{x}}_n,
	\label{E:alpha_1}
\end{equation}
and 
\begin{multline}
	\alpha_2 = \frac{(\dot{x}_n^{t_1-\delta})^2}{(\delta)^2} -
	\frac{\underline{\ddot{x}}_n}{\underline{\ddot{x}}_{n-1} (\delta)^2} (\dot{x}_{n-1}^{t_\K}+\underline{\ddot{x}}_{n-1}\theta)^2 + \\
	\frac{2\underline{\ddot{x}}_n}{(\delta)^2} \Big( x_{n-1}^{t_\K} + \dot{x}_{n-1}^{t_\K}\theta  + \frac{\underline{\ddot{x}}_{n-1}}{2}(\theta)^2 \\ - x_n^{t_1-\delta} - (\gamma+1)\dot{x}_n^{t_1-\delta}\delta - l_{n-1} - s_n \Big).
	\label{E:alpha_2}
\end{multline}
We denote the resulted feasible domain of \eqref{E:acceleration_2} by $\bm{\Lambda^{t_\srm}}$, with 
\begin{equation}
	\bm{\Lambda^{t_\srm}} = \big\{ \ddot{x}_n^{t_1} \big| \eqref{E:acceleration_2} \big\}.
\end{equation}
We name the safety constraint $\bm{\Lambda^{t_\srm}}$ as \emph{end-point constraint} as it considers the safety at the end-point of a potential hard brake. Note that $\bm{\Lambda^{t_\srm}}$ is similar to the safety constraint considered in the Gipps \cite{gipps1981behavioural} and RSS \cite{shalev2017formal} model, with the difference that $\bm{\Lambda^{t_\srm}}$ considers discrete instead of continuous signals.

\subsubsection{Acceleration domain at the midway-point of a potential hard brake}
\begin{figure}[h!] \vspace{-3mm}
	\centering
	\includegraphics[width=0.475\linewidth]{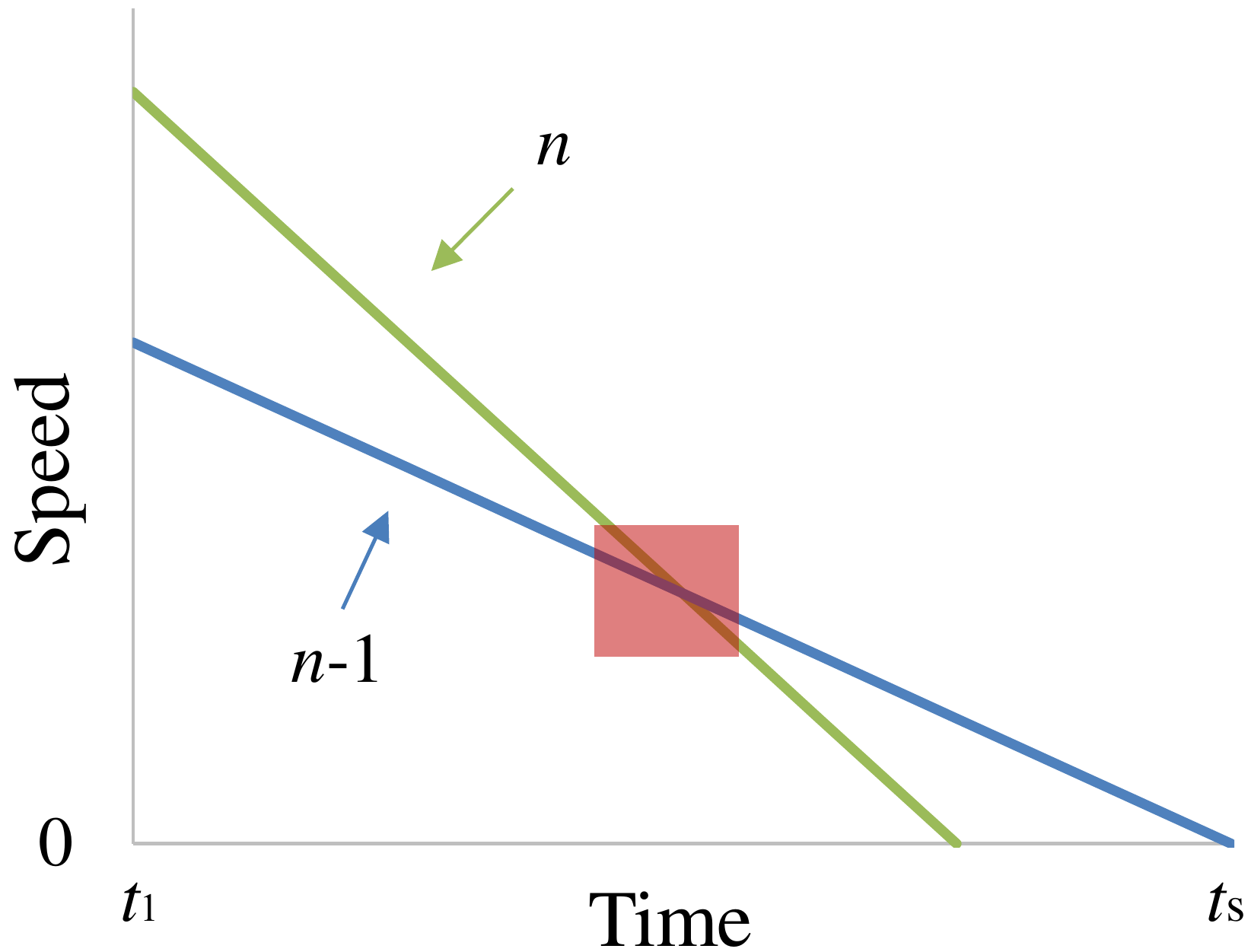}
	\caption{Hard-braking speed curves with existence of $t_\Lrm$ (\lightred{$\blacksquare$}: location of $t_\Lrm$).}
	\label{F:Spe_samples3}
\end{figure}

The 3rd constraint of \eqref{Ine:dist_c} works only when $t = t_\Lrm \in (t_1,t_\srm)$ exists. Fig. \ref{F:Spe_samples3} shows the only case when $t_\Lrm$ exists. Note that we exclude the scenario when the two speed curves overlap, i.e., $\F_n^{t}(t_\Lrm) = \F_n^{t}(t_1) = \F_n^{t}(t_s) = 0$. Obviously, there would be only one $t_\Lrm$ if the two speed curves are not overlapped. Hence, the only $t_\Lrm$ exists iff $\dot{x}_n^{t_1} > \dot{y}_{n-1}^{t_\K\rightarrow t_1}$, and $\dot{x}_n^{t_1}/\underline{\ddot{x}}_n > \dot{y}_{n-1}^{t_\K \rightarrow t_1}/\underline{\ddot{x}}_{n-1}$. Combined with \eqref{E:overline_s}, \eqref{E:B1} and \eqref{E:B2}, we have
\begin{align}
	\ddot{x}_n^{t_1} > & \frac{1}{\delta}\big( \dot{x}_{n-1}^{t_\K} + \underline{\ddot{x}}_{n-1} \theta - \dot{x}_n^{t_1-\delta} \big),
	\label{E:constraint_3a}\\
	\mathrm{and}\quad	\ddot{x}_n^{t_1} < & \frac{1}{\delta}\big( \frac{\underline{\ddot{x}}_n}{\underline{\ddot{x}}_{n-1}} (\dot{x}_{n-1}^{t_\K} + \underline{\ddot{x}}_{n-1} \theta) - \dot{x}_n^{t_1-\delta} \big).
	\label{E:constraint_3b}
\end{align}

Denote the resulted acceleration domain for existence of $t_\Lrm$ by $\bm{\widetilde{\Lambda}^{t_\Lrm}}$, with
	\begin{equation}
		\bm{\widetilde{\Lambda}^{t_\Lrm}} = \big\{ \ddot{x}_n^{t_1} \big| \eqref{E:constraint_3a},\eqref{E:constraint_3b} \big\}.
		\label{E:Lambda_b}
	\end{equation}
We have that i) if $\ddot{x}_n^{t_1} \not\in \bm{\widetilde{\Lambda}^{t_\Lrm}}$, 
the local maximum point $t_\Lrm \in (t_1,t_\srm)$ does not exist. ii) If $\ddot{x}_n^{t_1} \in \bm{\widetilde{\Lambda}^{t_\Lrm}}$, to find $t_\Lrm$, we let $\frac{\dd\mathcal{F}_n^{t_1}(t)}{\dd t}=0$. Hence, $t_\Lrm =  t_1 + \frac{\dot{x}_n^{t_1}-\dot{y}_{n-1}^{t_\K \rightarrow t_1}}{\underline{\ddot{x}}_{n-1} - \underline{\ddot{x}}_n}$ \footnote{Note that from \eqref{E:constraint_3a} and \eqref{E:constraint_3b}, we have $\underline{\ddot{x}}_n \neq \underline{\ddot{x}}_{n-1}$ for all $\ddot{x}_n^{t_1} \in \bm{\widetilde{\Lambda}^{t_\Lrm}}$.}, and $\F_n^{t}(t_\Lrm) = \frac{(\dot{x}_n^{t_1}-\dot{y}_{n-1}^{t_\K\rightarrow t_1})^2}{-2(\underline{\ddot{x}}_n-\underline{\ddot{x}}_{n-1})}$. Combined with \eqref{E:overline_s}, \eqref{E:B1} and \eqref{E:B2}, we have
\begin{equation}
	\frac{-\beta_1 -\sqrt{(\beta_1)^2 -4\beta_2}}{2}  \leq \ddot{x}_n^{t_1} \leq \frac{-\beta_1 +\sqrt{(\beta_1)^2 -4\beta_2}}{2},
	\label{E:acceleration_3}
\end{equation}
where
\begin{equation}
	\beta_1 = \frac{2}{\delta}\big( \dot{x}_n^{t_1-\delta} -\dot{x}_{n-1}^{t_\K} - \underline{\ddot{x}}_{n-1}\theta \big) + (2\gamma+1) (\underline{\ddot{x}}_{n-1} - \underline{\ddot{x}}_n),
	\label{E:beta_1}
\end{equation}
and 
\begin{multline}
	\beta_2 = \frac{1}{(\delta)^2} \big( \dot{x}_{n-1}^{t_\K} + \underline{\ddot{x}}_{n-1} \theta - \dot{x}_n^{t_1-\delta} \big)^2 -\\ 
	\frac{2(\underline{\ddot{x}}_{n-1} - \underline{\ddot{x}}_n)} {(\delta)^2} \Big( x_{n-1}^{t_\K} + \dot{x}_{n-1}^{t_\K}\theta  + \frac{\underline{\ddot{x}}_{n-1}}{2}(\theta)^2 \\ - x_n^{t_1-\delta} - (\gamma+1)\dot{x}_n^{t_1-\delta}\delta - l_{n-1} - s_n \Big).
	\label{E:beta_2}
\end{multline}
We denote the resulted feasible acceleration domain of the 3rd constraint in \eqref{Ine:dist_c} by $\bm{\Lambda^{t_\Lrm}}$, with
\begin{equation}
	\bm{\Lambda^{t_\Lrm}} = \big\{ \ddot{x}_n^{t_1} \big| \eqref{E:constraint_3a},\eqref{E:constraint_3b}, \eqref{E:acceleration_3}\big\}.
\end{equation}
We name the safety constraint $\bm{\Lambda^{t_\Lrm}}$ as \emph{midway-point constraint} as it considers the safety at the midway-point of a potential hard brake. Note that $\bm{\Lambda^{t_\Lrm}}$ is also considered in some recent SOCF models for CAVs. Our considerations differ from theirs in the way that we incorporate the influence of discrete signals.

\subsubsection{Acceleration determination}
Once safety is guaranteed, we assume that vehicles tend to maximize the car-following efficiency. That is, vehicles always choose the maximum acceleration within the feasible domain of safety. At every decision moment, we have
\begin{equation}
	\ddot{x}_n^{t_1} = 
	\begin{cases} 
		a_\mathrm{a}, & \mbox{if } 
		a_\mathrm{a} \not\in \bm{\widetilde{\Lambda}^{t_\Lrm}}\\
		a_\mathrm{b}, & \mbox{if } 
		a_\mathrm{a} \in \bm{\widetilde{\Lambda}^{t_\Lrm}} 
	\end{cases}. 
	\label{E:ddotxn}
\end{equation}
where
\begin{align}
	a_\mathrm{a} = & \max\big\{\bm{\Lambda^0} \cap \bm{\Lambda^{t_1}} \cap \bm{\Lambda^{t_\srm}} \big\}, \label{E:alpha_b}\\
	a_\mathrm{b} = & \max\big\{\bm{\Lambda^0} \cap \bm{\Lambda^{t_1}} \cap \bm{\Lambda^{t_\srm}} \cap \bm{\Lambda^{t_\Lrm}} \big\}. \label{E:alpha_c}	
\end{align}

\section{SIMULATION}
In this section, we will first test the impact of \eqref{Ine:dist_c}'s each safety constraint on the car-following safety. After that, we will investigate the safety, efficiency, and string stability of the proposed SOCF model, and then test the influence of packet loss on the model's performance.

\subsection{Simulation Settings}
We set three types of CAVs in our simulation, including small vehicle, midsize vehicle, and large vehicle. They are different in length, acceleration performance, and mechanical delay, as shown in Table \ref{T:dif_veh}. It is worth mentioning that even for a fixed vehicle type, the mechanical delay may vary depending on the executed acceleration in practice. We set a fixed mechanical delay for each vehicle type in the simulation just for simplicity. A dynamic mechanical delay can be easily incorporated into the model framework. We adopt the data in \cite{maurya2012study} for setting the minimum accelerations, and the data in \cite{rajamani2011vehicle, maurya2012study, xiao2011practical, ploeg2013lp} for setting the mechanical delays. Note that large vehicles, such as trucks and buses, usually use air brakes, which makes them have much larger mechanical delays than other vehicles using hydraulic brakes.
\begin{table}[h!]
	\caption{Parameters for different vehicle types}
	\label{T:dif_veh}
	\begin{tabular}{cc|clll}
		\hline
		\multicolumn{2}{c|}{Vehicles}&\multicolumn{4}{c}{Coefficients}                                                                                                                 \\
		Types           & Examples       &\multicolumn{1}{l}{$l$ ($\mathrm{m}$)} & $\overline{\ddot{x}}$ ($\mathrm{m/s}^2$) & $\underline{\ddot{x}}$ ($\mathrm{m/s}^2$) & $\epsilon$ ($\mathrm{s}$)\\ \hline
		small veh   & \tabincell{c}{compact car, \\ midsize car\\} &   4.5   &                 ~~~~~1                        &                ~~-1.5                       &    0.07          \\
		midsize veh & \tabincell{c}{minibus, \\ pickup truck\\}    &   7.5   &                 ~~~~0.9                      &                ~~-0.9                       &    0.15          \\
		large veh   & \tabincell{c}{bus, \\truck\\}                &   15    &                 ~~~~0.6                      &                ~~-0.6                       &    ~0.5           \\ \hline
	\end{tabular}
\end{table}

Table \ref{T:coe} lists the values we used in the simulation for other main coefficients. Specifically, the communication interval $\delta$ is set to 0.1 s according to a commonly used standard by Dedicated Short Range Communication (DSRC) \cite{kenney2011dedicated}. Some researchers have pointed out that the total communication delay should not exceed 0.1 s for close-range cooperative platooning \cite{reichardt2002cartalk}. Existing communication protocols are able to control the delay within 0.02 s\cite{tokuda2000dolphin, durresi2005emergency}. In this paper, we assume that the transmission delay $\tau$ follows a uniform distribution between 0.04 and 0.08 s. $\underline{\kappa}$ is dynamically determined by $\tau$. To \red{improve the driving comfort} and reduce acceleration fluctuations due to the randomness in transmission delays and phase difference, we let each CAV store all $\underline{\kappa}$ within the last 10 s, and take the maximum $\underline{\kappa}$ as its real-time $\kappa$. The phase difference for each PV-FV pair is initialized randomly at the beginning of the simulation. After initialization, $\phi$ is fixed during the whole simulation process. According to a real-world reliability evaluation of IEEE 802.11 p-based vehicle-to-vehicle communication, the packet loss rate can be as high as 30\% \cite{wang2015reliability}, we hence test five kinds of packet loss rates with the maximum value of 50\% to represent an extremely poor communication environment.
\begin{table}[h!]
	\caption{Value for main coefficients}
	\label{T:coe}
	\begin{tabular}{c|l}
		\hline
		Coefficients                                   	 & Values in simulation           \\ \hline
		information/decision interval $\delta$       	 & $0.1 ~\mathrm{s}$    			\\
		transmission delay $\tau$ (dynamic)     			 & from $0.04$ to $0.08 ~\mathrm{s} $   	\\		
		decision duration $d$							 & $0$ \\
		phase difference $\phi$    			 			 & from $0$ to $0.1$ ~$\mathrm{s} $  		\\
		packet loss rate               	 	 			 &  $0\%$, $1\%$, $10\%$, $25\%$ or $50\%$ 	\\          		
		stop gap $s$                                	 &  $1 ~\mathrm{m}$   \\
		max speed $\overline{\dot{x}}$         		     & from $8$ to $22 ~\mathrm{m/s} $   \\
		additional gap coefficient $\gamma$				 &  $5$  \\      \hline
	\end{tabular} \vspace{-3mm}
\end{table}

\subsection{Influence of Safety Constraints}
As mentioned in Sec. \ref{S:Intro}, the SOCF models in the literature either only consider the end-point constraint or the midway-point \& end-point constraints, while some efficiency-oriented or stability-oriented car-following models only consider the start-point constraint. To the best of our knowledge, few studies have integrated all three constraints, especially in the discrete signal system framework. The lack of some constraint(s) does not necessarily lead to crashes in daily driving, but will suddenly cause a tragedy under some particular conditions. To investigate the influence of different safety constraints ($\bm{\Lambda^{t_1}}$, $\bm{\Lambda^{t_\srm}}$ and $\bm{\Lambda^{t_\Lrm}}$) on the car-following safety, we design three special scenarios. In these scenarios, we remove each safety constraint respectively and see how a crash happens. Note that without a necessary constraint, there may be no feasible acceleration in some conditions, which implies a crash risk. Once this happens, the vehicle will brake hard until it finds a feasible solution.  

\subsubsection{Without the start-point constraint (let $\bm{\Lambda^{t_1}} \equiv \RR$)}
\begin{figure}[h!]
	\centering
	\subfloat[][Without the start-point constraint]{\resizebox{0.245\textwidth}{!}{
			\includegraphics[width=0.2\textwidth]{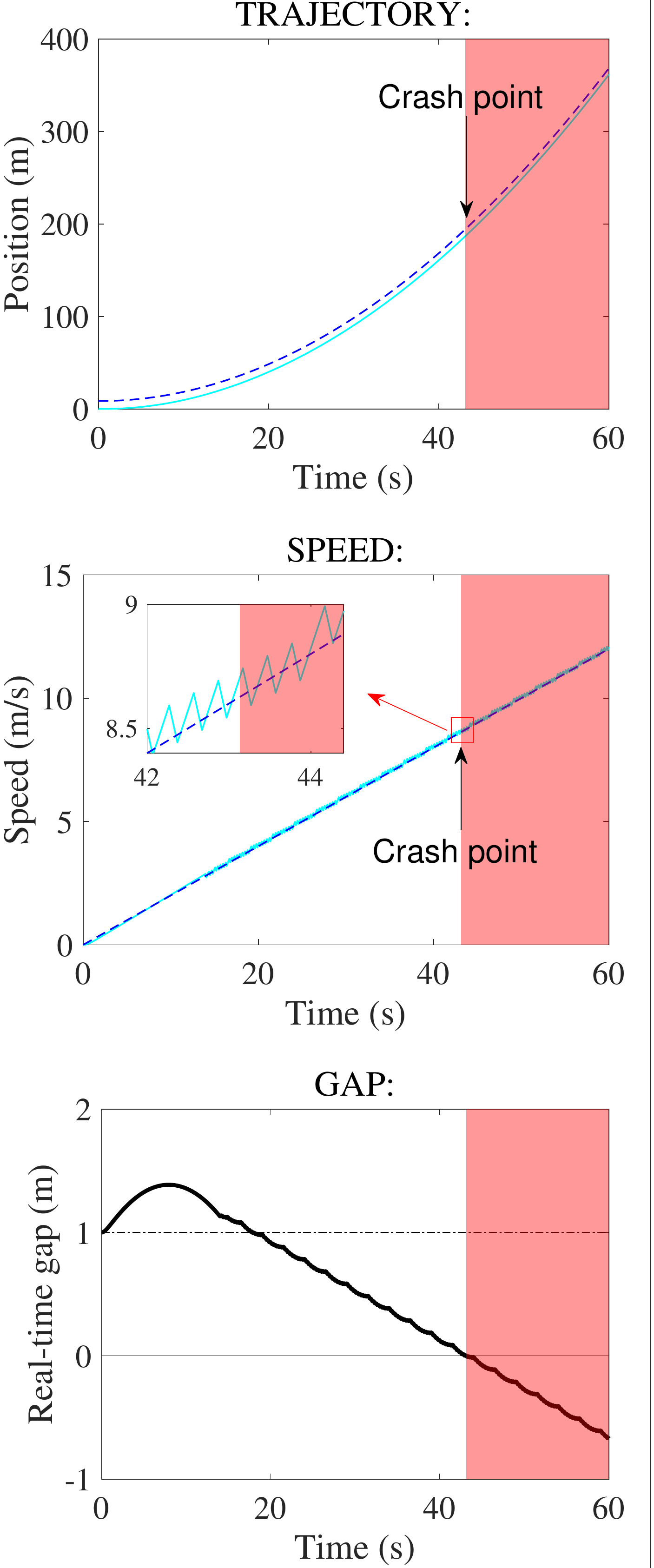}}
		\label{fig:3_1_1}
	}
	\subfloat[][With complete constraints ]{\resizebox{0.245\textwidth}{!}{
			\includegraphics[width=0.2\textwidth]{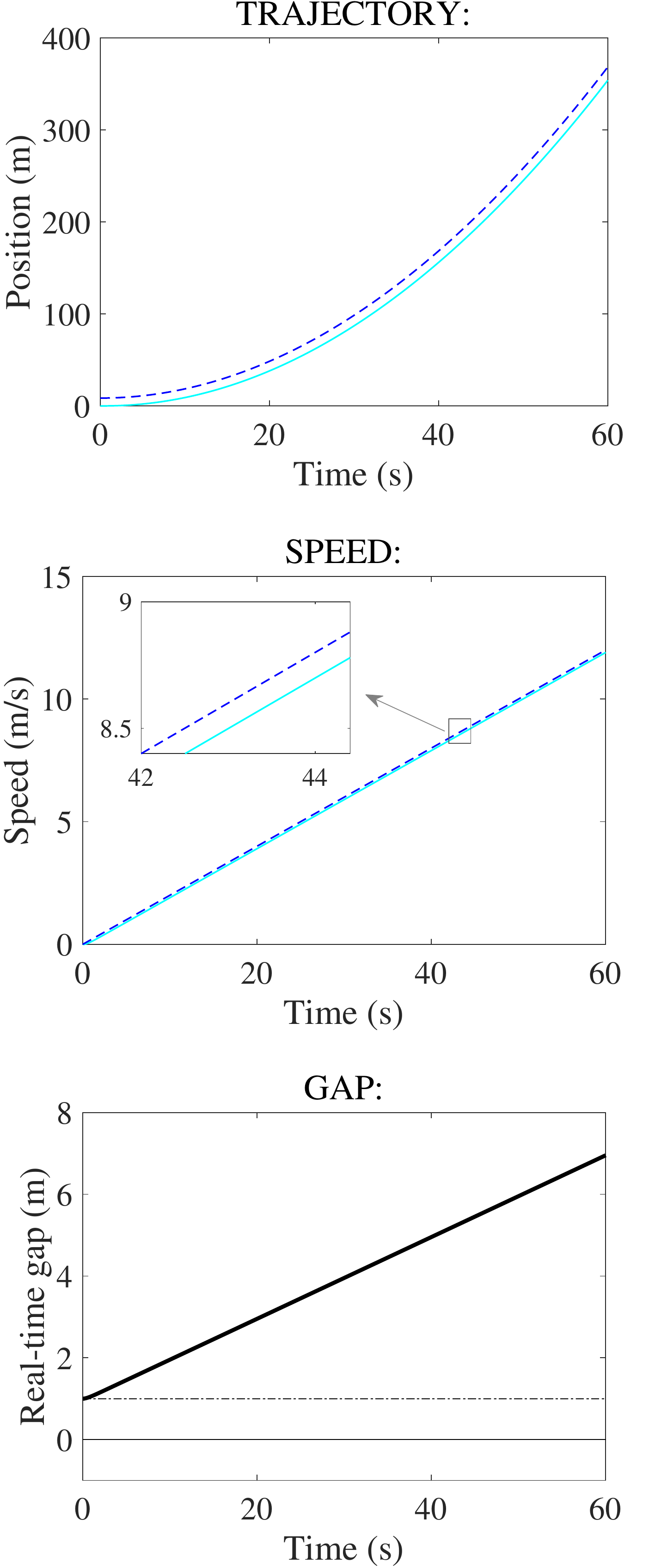}}
		\label{fig:3_1_2}
	}
	
	\subfloat{\resizebox{0.3\textwidth}{!}{
			\includegraphics[width=1\textwidth]{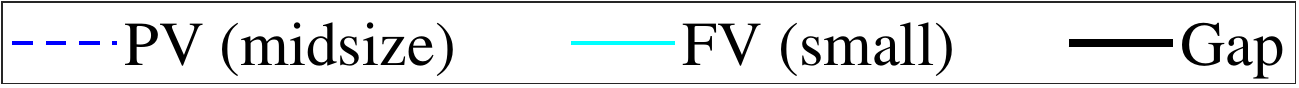}}
		\label{fig:3_1_legend}}
	\caption{Comparison between with and without the start-point constraint.}
	\label{F:3_1}
\end{figure}
By making $\bm{\Lambda^{t_1}} \equiv \RR$ in \eqref{E:ddotxn}-\eqref{E:alpha_c}, we can relax the start-point constraint. In this scenario, we consider a small CAV following a midsize CAV. The initial bumper-to-bumper gap is 1 m. Both vehicles have a speed of 0 in the beginning, and the PV accelerates with a constant acceleration of 0.2 $\mathrm{m/s}^2$. Fig. \ref{F:3_1} shows the FV following the PV according to \eqref{E:ddotxn}-\eqref{E:alpha_c}, with $\bm{\Lambda^{t_1}}$ relaxed or 
retained, respectively. Fig. \ref{F:3_1} \subref{fig:3_1_1} shows the result when $\bm{\Lambda^{t_1}}$ is relaxed, while Fig. \ref{F:3_1} \subref{fig:3_1_2} shows the result with complete constraints for comparison. Note that to see clearly how the model influences the car-following behavior, we allow the FV to ``pass through" the PV after the crash point as if the crash does not happen (as shown in the red area of the figure). That is why we will see negative gaps in the figure \footnote{In case of a negative gap, the FV is still considered to be following the PV and behaves according to the car-following model.}. 

As can be seen in Fig. \ref{F:3_1} \subref{fig:3_1_1}, when $\bm{\Lambda^{t_1}}$ is relaxed, after a short period of normal driving, the FV will approach the PV gradually, and finally collide with it. If we focus on the speed curves and zoom in, we could see how the crash happens. In Fig. \ref{F:3_1} \subref{fig:3_1_1}, without the start-point constraint, the FV's speed keeps zigzagging around the PV's speed (with a higher probability of being faster than the PV's speed); in Fig. \ref{F:3_1} \subref{fig:3_1_2}, with the complete constraints, the FV's speed keeps slightly slower than the PV's speed. As a result, the gap in Fig. \ref{F:3_1} \subref{fig:3_1_2} keeps increasing, while the gap in Fig. \ref{F:3_1} \subref{fig:3_1_1} keeps fluctuating with a decreasing tendency in general, and finally leads to a rear-end crash.

\begin{figure}[h!]
	\centering
	\includegraphics[width=0.65\linewidth]{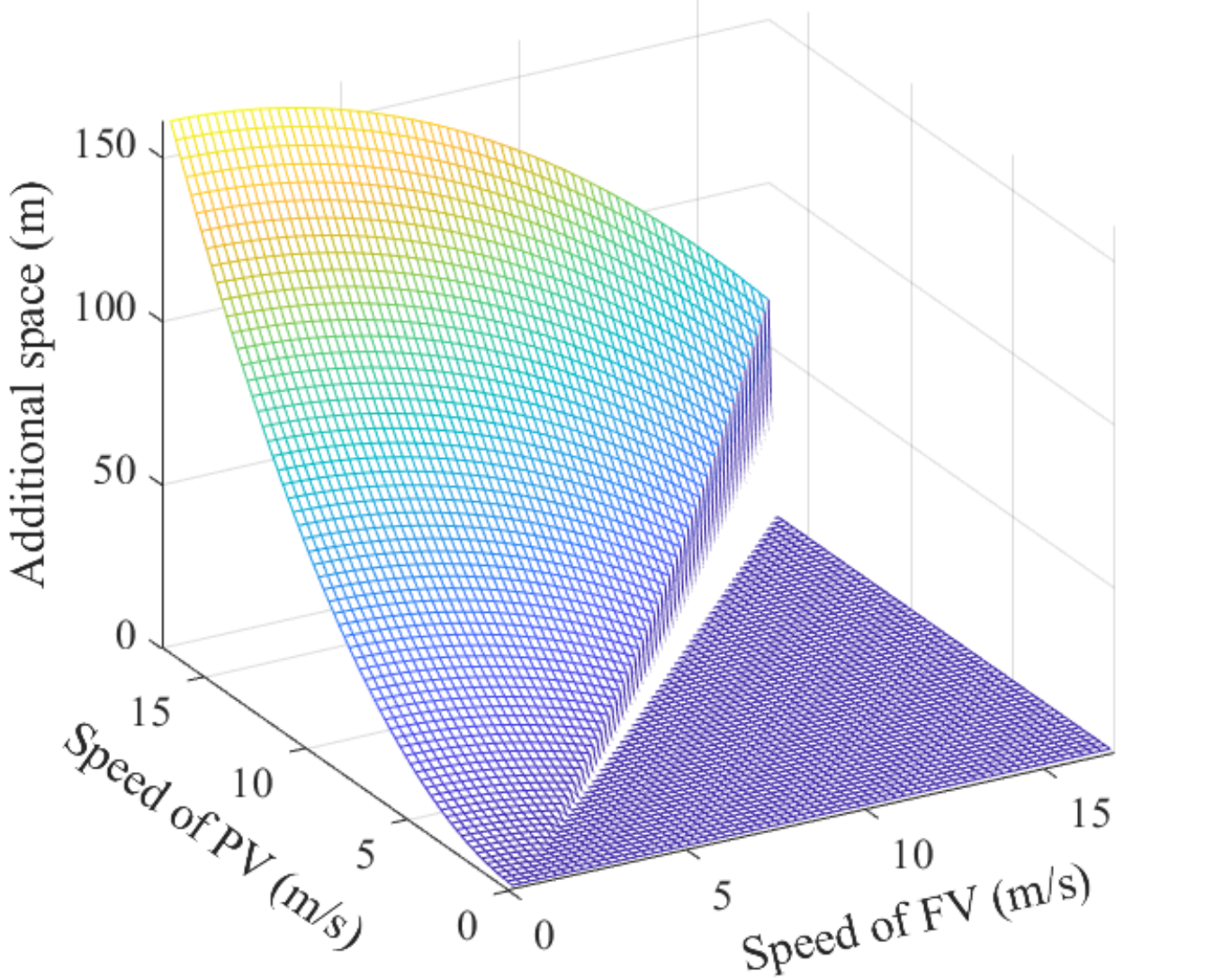}
	\caption{Additional gap between a midsize PV and a small FV caused by the start-point constraint.}
	\label{fig:addispaceno1}
\end{figure}

To see why the zigzags in the FV's speed happen, we show the additional gap caused by $\bm{\Lambda^{t_1}}$ under different speed pairs of a midsize PV and a small FV in Fig. \ref{fig:addispaceno1}. We first calculate the minimum car-following distance considering all safety constraints and then calculate the minimum car-following distance in the absence of $\bm{\Lambda^{t_1}}$. The additional gap is then obtained by subtracting the second distance from the first distance. 

From Fig. \ref{fig:addispaceno1}, we can see that the influence of the start-point constraint mainly lies in the area when the PV's speed is not lower than the FV's speed, and there is a big ``jump'' when two vehicles have the same speed. Hence when the FV has a slower speed than the PV, the start-point constraint takes effect, otherwise, other constraints take effect. That is the reason why speed's zigzags happen without the start-point constraint: when the FV's speed is lower than the PV's speed, the FV is suggested to accelerate because it mistakenly believes that the gap is more than enough due to the absence of $\bm{\Lambda^{t_1}}$. Once the FV's speed exceeds the PV's speed, other safety constraints begin to take effect and the FV suddenly realizes that it has to decelerate. This process keeps repeating and results in zigzags in the speed. 

We note that the collision in Fig. \ref{F:3_1} \subref{fig:3_1_1} is unique to a system with discrete signals. This is because vehicles do not need to wait until the next decision-making moment to react in a continuous signal system, and hence the zigzags along with the resulted crash can be avoided. 

\subsubsection{Without the end-point constraint (let $\bm{\Lambda^{t_\srm}} \equiv \RR$)}
By making $\bm{\Lambda^{t_\srm}} \equiv \RR$ in \eqref{E:ddotxn}-\eqref{E:alpha_c}, we can relax the end-point constraint.
\begin{figure}[h!]
	\centering
	\subfloat[][Without the end-point constraint]{\resizebox{0.245\textwidth}{!}{
			\includegraphics[width=0.2\textwidth]{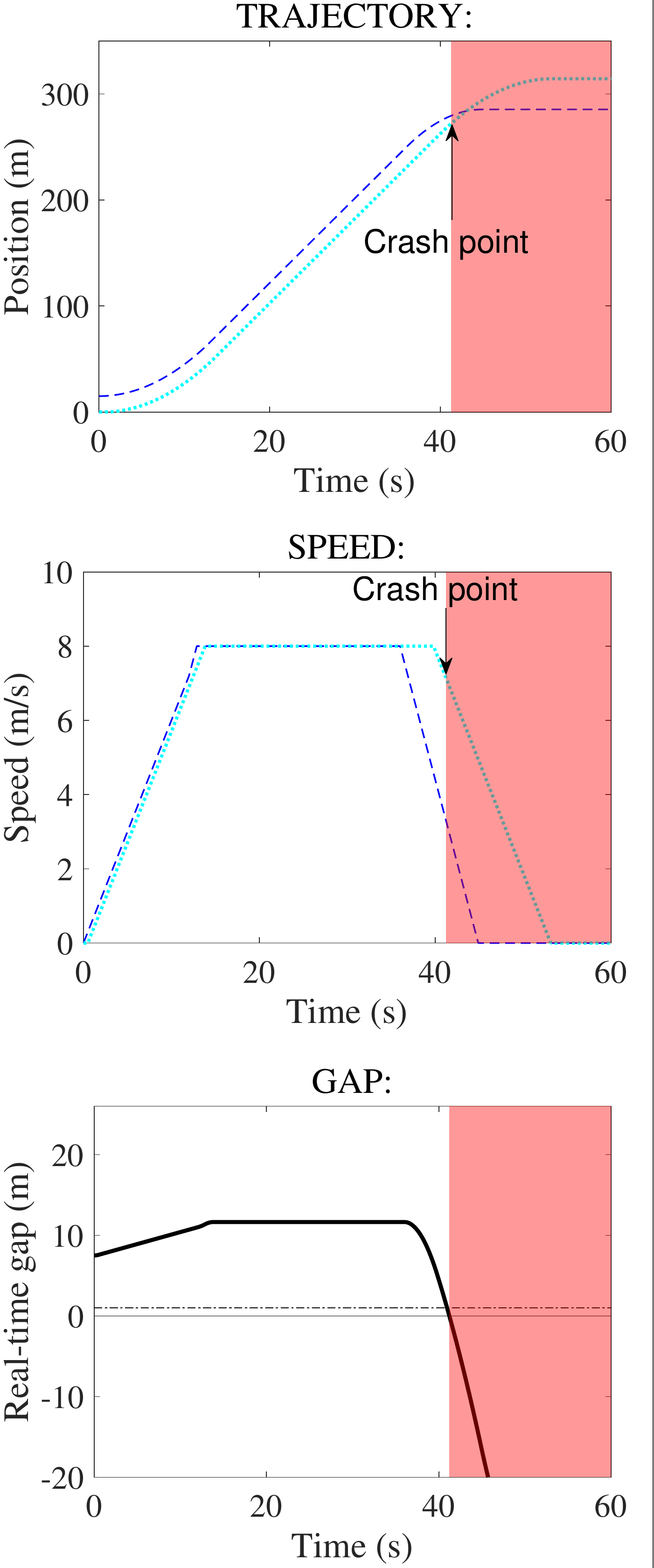}}
		\label{fig:3_2_1}
	}
	\subfloat[][With complete constraints]{\resizebox{0.245\textwidth}{!}{
			\includegraphics[width=0.2\textwidth]{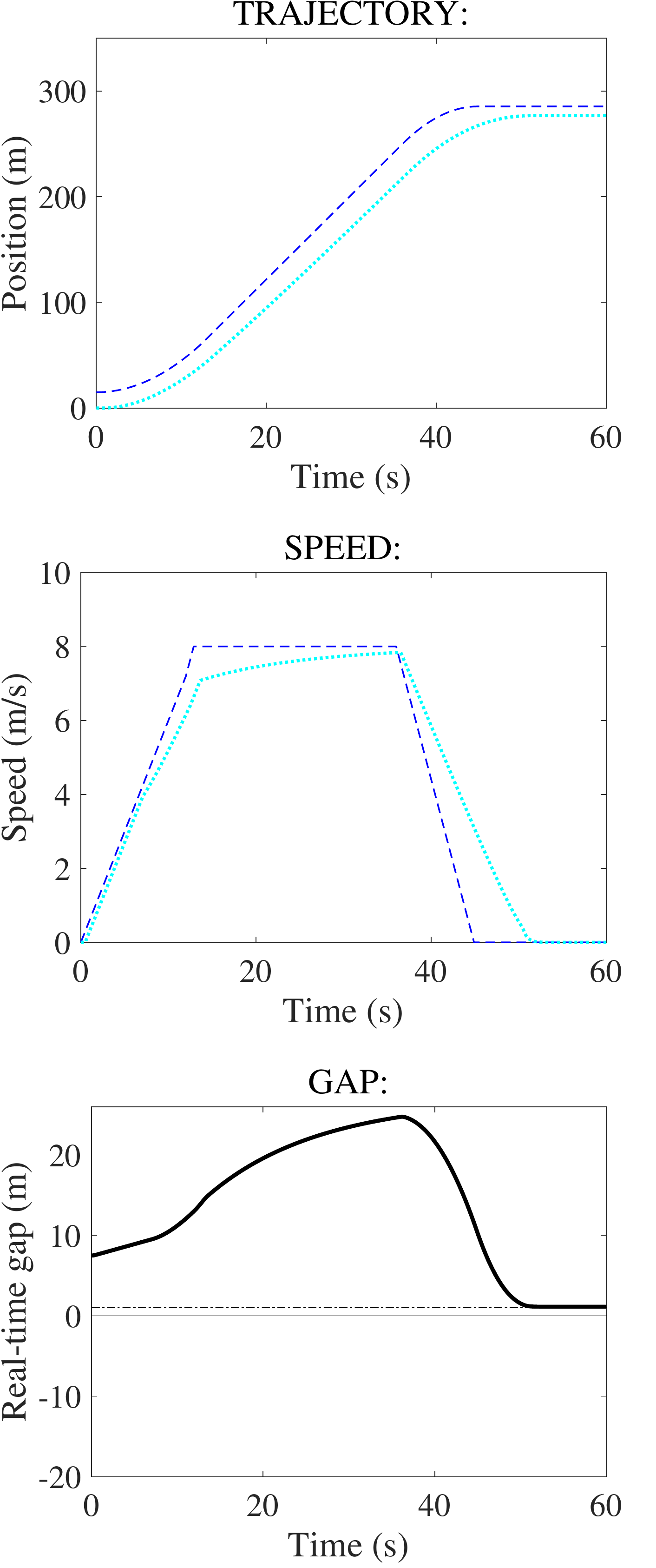}}
		\label{fig:3_2_2}
	}
	
	\subfloat{\resizebox{0.3\textwidth}{!}{
			\includegraphics[width=1\textwidth]{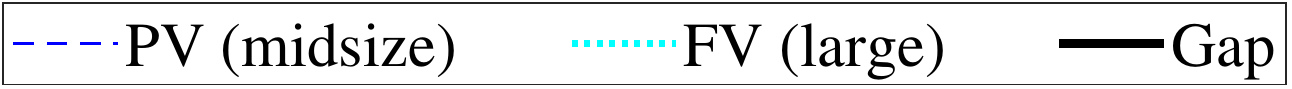}}
		\label{fig:3_2_legend}}
	\caption{Comparison between with and without the end-point constraint.}
	\label{F:3_2}
\end{figure}
In this scenario, we consider a large CAV following a midsize CAV. The initial bumper-to-bumper gap is 7.5 m, and the maximum speed for both vehicles is around 30 km/h. Both vehicles have a speed of 0 in the beginning. The PV accelerates from 0 to the maximum speed, and then travels with the maximum speed until a sudden hard brake. Fig. \ref{F:3_2} shows the FV following the PV according to \eqref{E:ddotxn}-\eqref{E:alpha_c}, with $\bm{\Lambda^{t_\srm}}$ relaxed or retained, respectively. Fig. \ref{F:3_2} \subref{fig:3_2_1} shows the result when $\bm{\Lambda^{t_\srm}}$ is relaxed, while Fig. \ref{F:3_2} \subref{fig:3_2_2} shows the result with complete constraints for comparison. 

\begin{figure}[h!]
	\centering
	\includegraphics[width=0.65\linewidth]{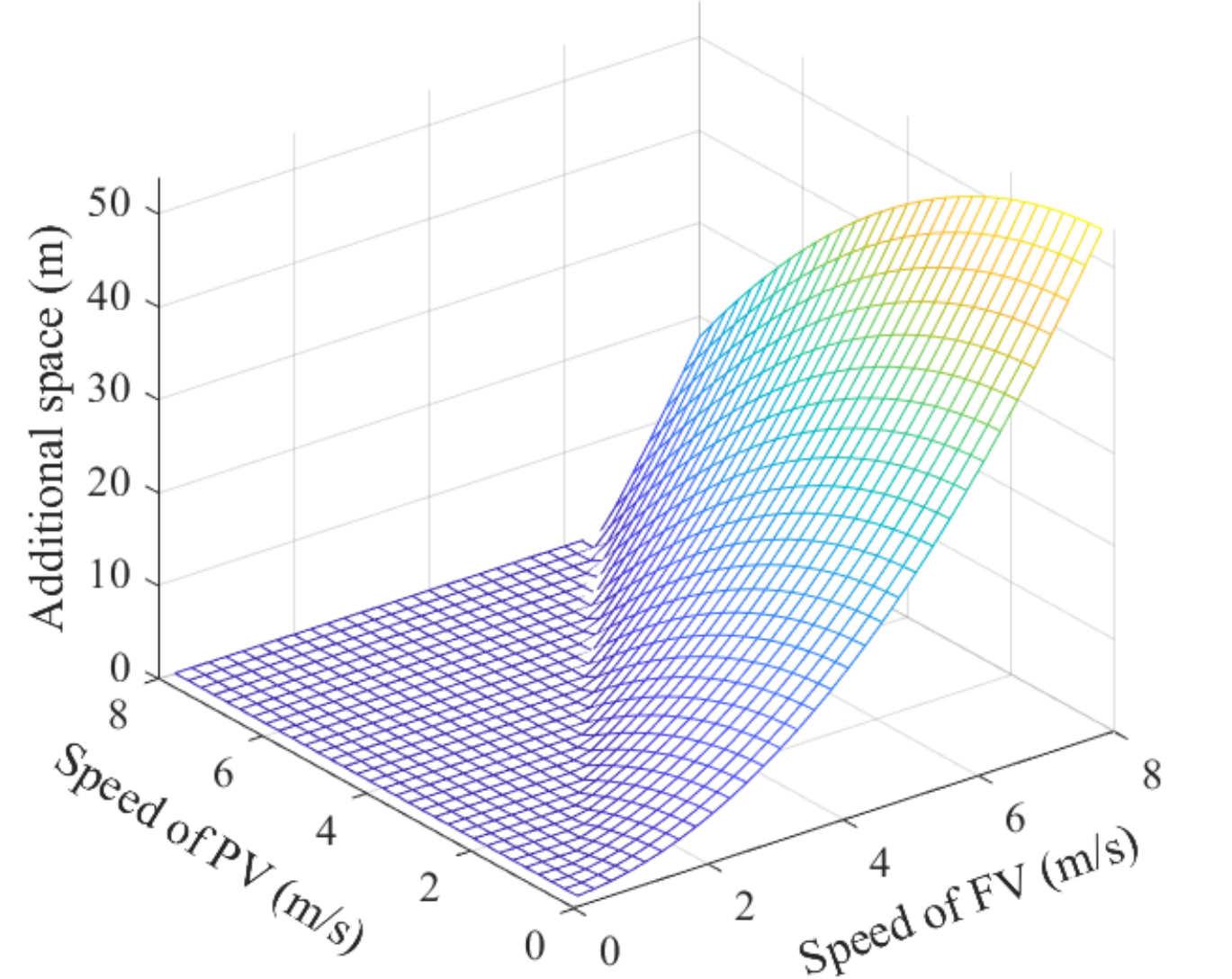}
	\caption{Additional gap between a midsize PV and a large FV caused by the end-point constraint.}
	\label{fig:addispaceno2}
\end{figure}

We can see from Fig. \ref{F:3_2} \subref{fig:3_2_1} that a crash happens during the hard brake of two vehicles without the end-point constraint. The sudden hard brake of PV starts when both vehicles form a stable car-following status with a constant gap. Compared with the complete constraints in Fig. \ref{F:3_2} \subref{fig:3_2_2}, we see that the FV's trajectory starts to deviate much earlier (before 10 s), implying that the crash risk has existed for a long time. When both vehicles have a speed of around 8 m/s, the FV in Fig. \ref{F:3_2} \subref{fig:3_2_2} keeps an additional gap of about 20 m compared with the FV in Fig. \ref{F:3_2} \subref{fig:3_2_1}. Fig. \ref{fig:addispaceno2} shows the complete map of the additional gap caused by $\bm{\Lambda^{t_\srm}}$ under different speed pairs of a midsize PV and a large FV. We can see that $\bm{\Lambda^{t_\srm}}$ mainly takes effect when the PV's speed is lower than the FV's speed. That is why after a hard brake of the PV, the FV in Fig. \ref{F:3_2} \subref{fig:3_2_1} reacts too late to avoid a rear-end crash: the end-point constraint is lacking, hence the FV does not realize that it is too close to the PV.

\subsubsection{Without the midway-point constraint (let $\bm{\Lambda^{t_\Lrm}} \equiv \RR)$}
\begin{figure}[h!]
	\centering
	\subfloat[][Without the midway-point constraint]{\resizebox{0.245\textwidth}{!}{
			\includegraphics[width=0.2\textwidth]{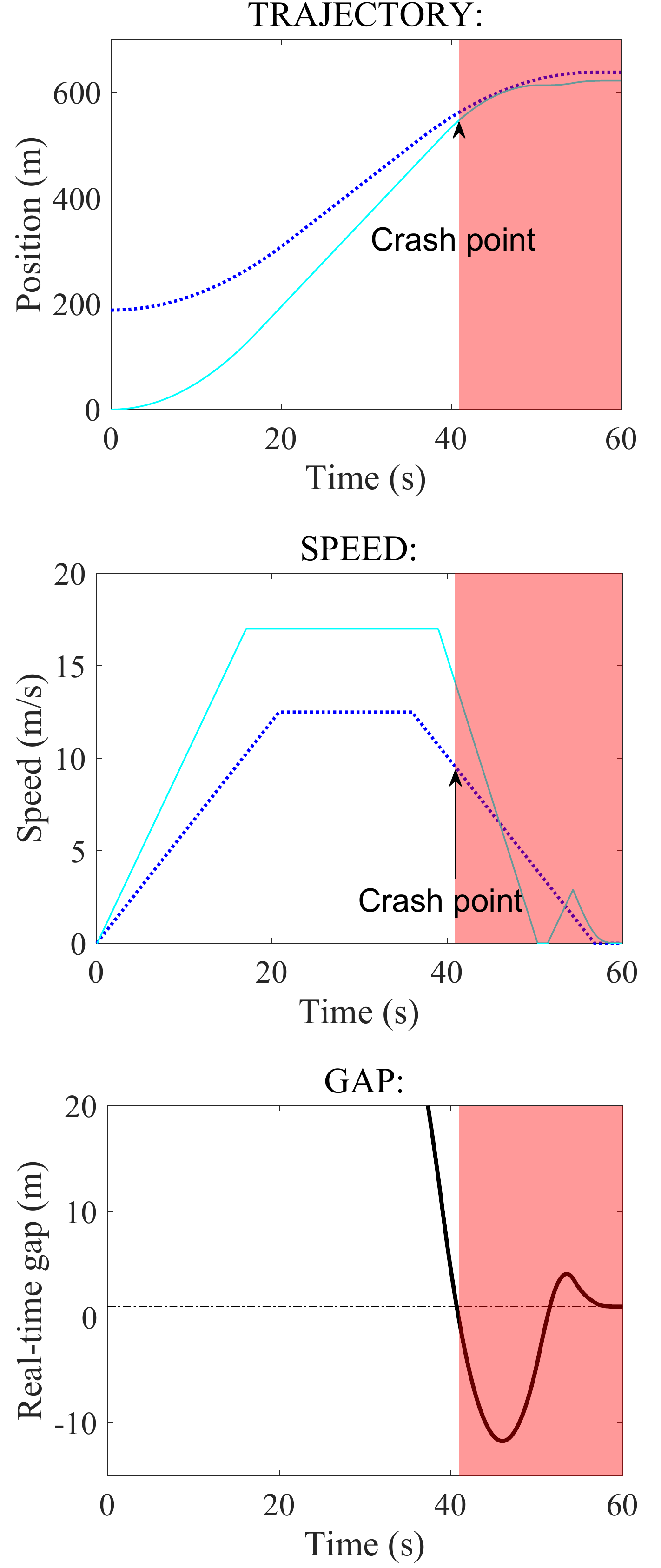}}
		\label{fig:3_3_1}
	}
	\subfloat[][With complete constraints]{\resizebox{0.245\textwidth}{!}{
			\includegraphics[width=0.2\textwidth]{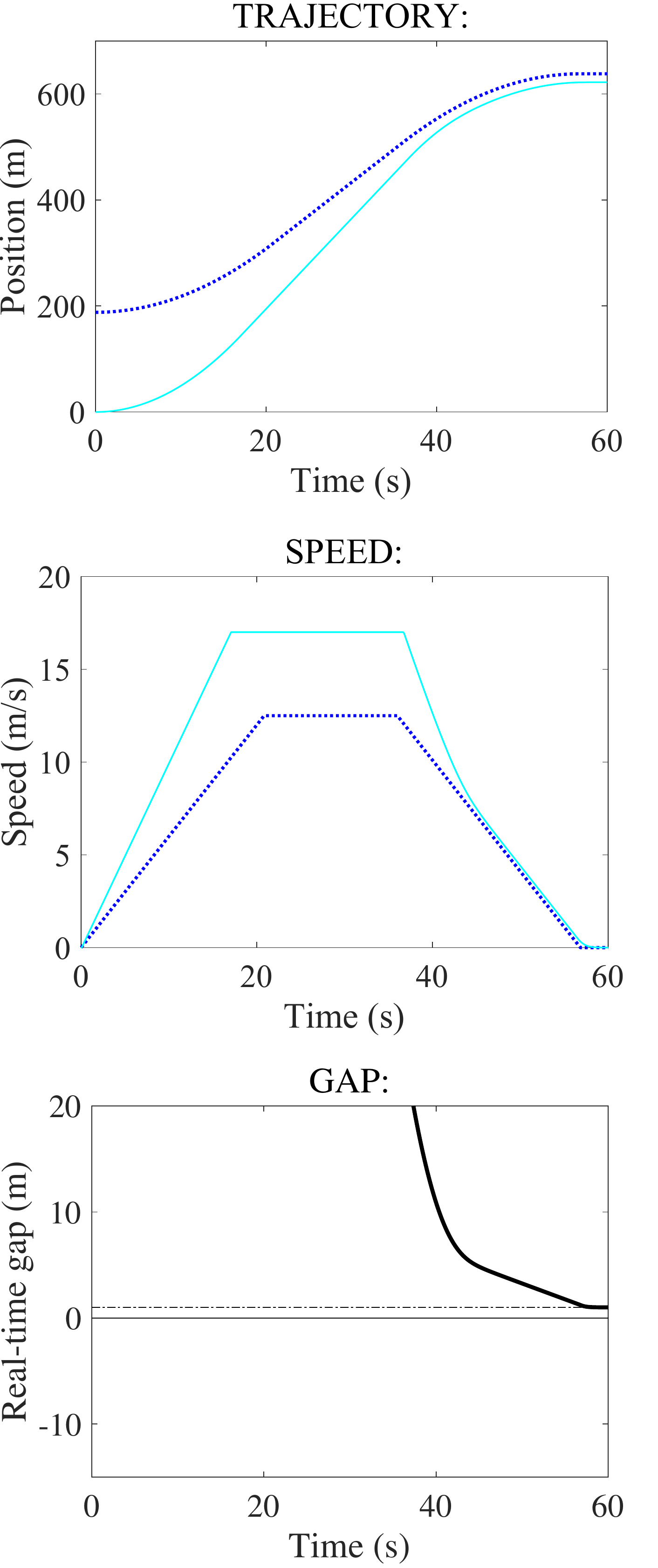}}
		\label{fig:3_3_2}
	}
	
	\subfloat{\resizebox{0.3\textwidth}{!}{
			\includegraphics[width=1\textwidth]{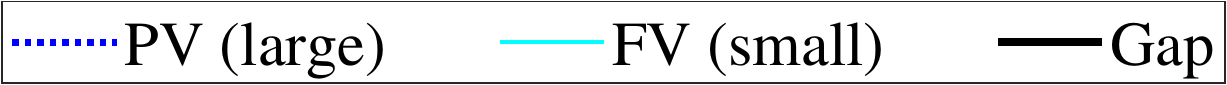}}
		\label{fig:3_3_legend}}
	\caption{Comparison between with and without the midway-point constraint.}
	\label{F:3_3}
\end{figure}
By making $\bm{\Lambda^{t_\Lrm}} \equiv \RR$ in \eqref{E:ddotxn}-\eqref{E:alpha_c}, we can relax the midway-point constraint. In this scenario, we consider a small CAV following a large CAV.
The initial bumper-to-bumper gap is 173 m, the maximum speed for the large PV is around 45 km/h, and the maximum speed for the small FV is around 60 km/h. Both vehicles have a speed of 0 in the beginning. The PV accelerates from 0 to the maximum speed and then travels with the maximum speed until a sudden hard brake. Fig. \ref{F:3_3} shows the FV following the PV according to \eqref{E:ddotxn}-\eqref{E:alpha_c}, with $\bm{\Lambda^{t_\Lrm}}$ relaxed or retained, respectively. Fig. \ref{F:3_3} \subref{fig:3_3_1} shows the result when $\bm{\Lambda^{t_\Lrm}}$ is relaxed, while Fig. \ref{F:3_3} \subref{fig:3_3_2} shows the result with complete constraints for comparison.

We can see from Fig. \ref{F:3_3} \subref{fig:3_3_1} that a crash happens during the hard brake of two vehicles without the midway-point constraint. The hard brake of PV starts when the FV is still chasing the PV (a stable gap has not been reached). Similar to the last scenario, the FV reacts too late after a hard brake of PV to avoid a rear-end crash when the midway-point constraint is lacking. It is worth mentioning that, if we look at the time when both vehicles come to a stop in Fig. \ref{F:3_3} \subref{fig:3_3_1}, we will find that the gap is equal to the stopgap we set in Table \ref{T:coe}, which is 1 m. That is, safety is ``guaranteed'' when both vehicles come to a full stop. However, the rear-end collision has already happened before the stop. Therefore, just thinking about safety at the stop, like models in most literature (e.g., \cite{gipps1981behavioural,van2006impact}), will lead to a crash in this scenario.

\begin{figure}[h!]
	\centering
	\includegraphics[width=0.65\linewidth]{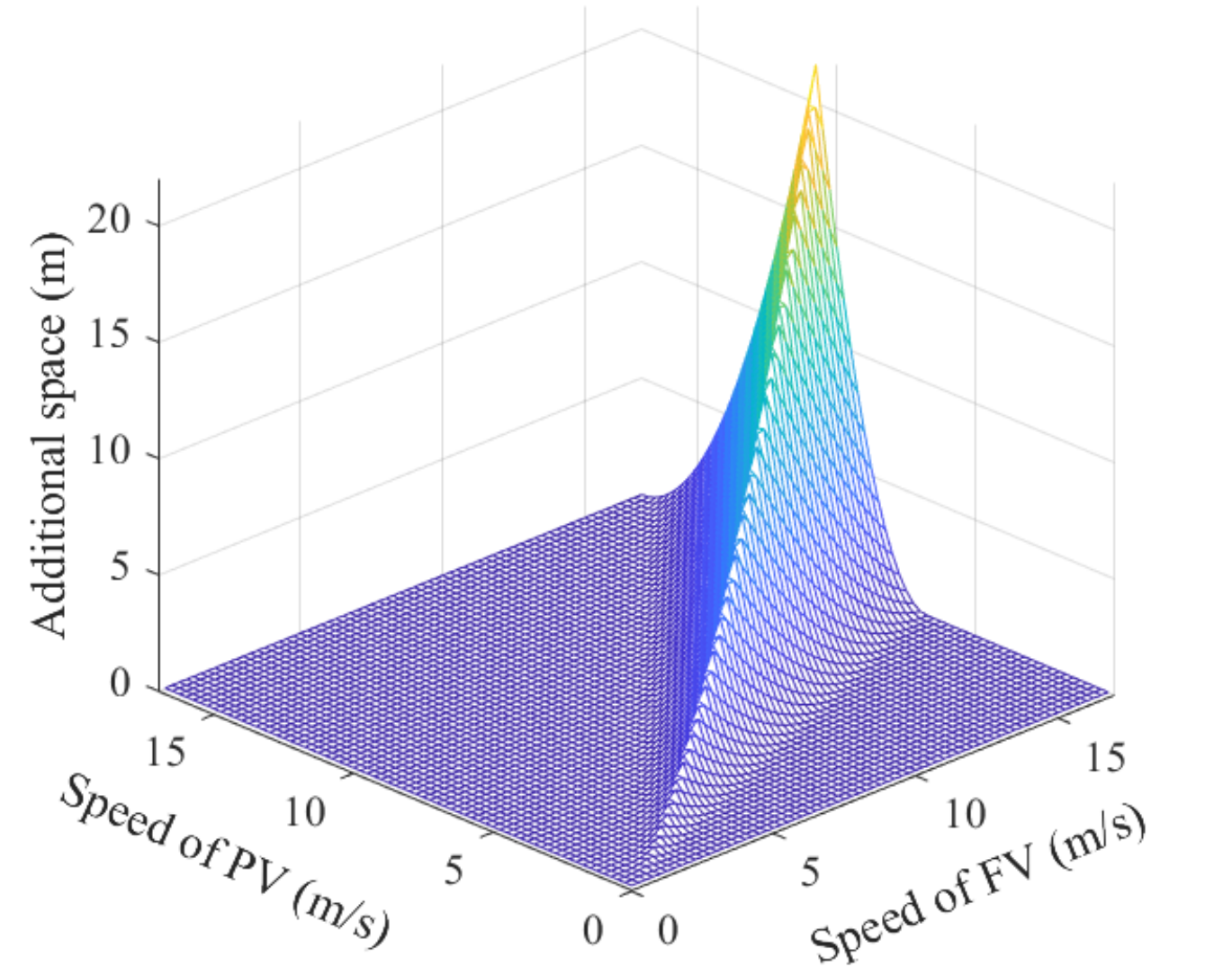}
	\caption{Additional gap between a large PV and a small FV caused by the midway-point constraint.}
	\label{fig:addispaceno3}
\end{figure}

Fig. \ref{fig:addispaceno3} shows the complete map of the additional gap caused by $\bm{\Lambda^{t_\Lrm}}$ under different speed pairs of a large PV and a small FV. We can see that $\bm{\Lambda^{t_\Lrm}}$ mainly takes effect when the FV's speed is a bit faster than the PV's speed. Hence, in a short period just after the hard brake of the PV in Fig. \ref{F:3_3} \subref{fig:3_3_1}, the PV's speed is just a bit lower than the FV's speed, hence the FV is unaware of the danger due to absence of $\bm{\Lambda^{t_\Lrm}}$. When the PV's speed decreases further and other safety constraints begin to take effect, the FV realizes that the gap is too small and begins to brake hard. However, it is already too late and the rear-end crash is unavoidable. 

\subsubsection{Summary}
As can be seen from the three collision examples, a car-following model that lacks any safety constraint is just like a system with a bug: although the bug can only be triggered in specific conditions, once such conditions happen, the bug will lead to system paralysis. Specifically, lack of the start-point constraint could lead to a collision during the normal driving process, while lack of the midway-point or end-point constraint could lead to a collision in case of an emergency hard brake. With the integration of all three constraints in one model, our proposed model can ensure car-following safety during the whole driving process of CAVs with discrete signals.

\subsection{Efficiency analysis}
\begin{figure}[h!]
	\centering
	\includegraphics[width=0.65\linewidth]{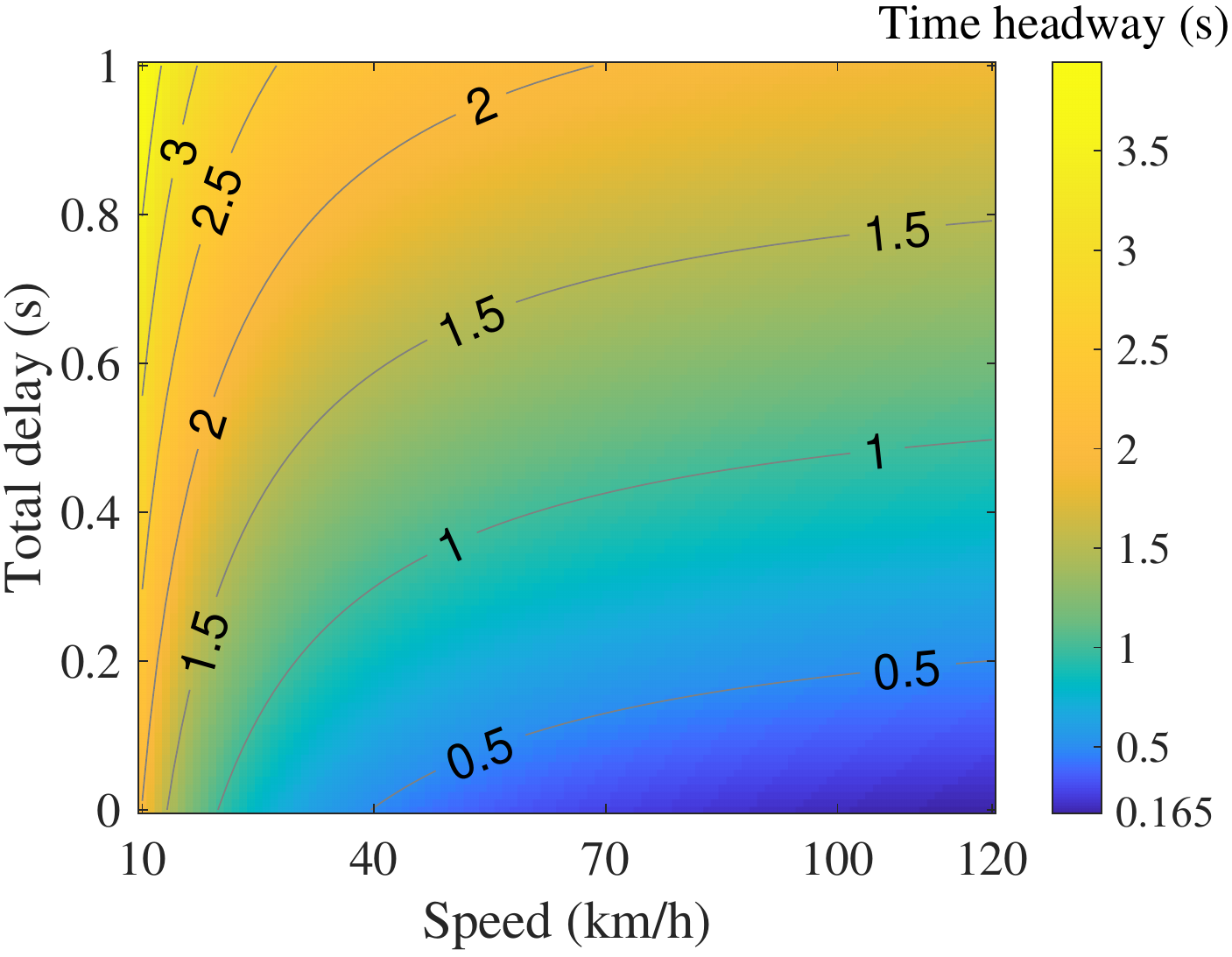}
	\caption{Influence of total delay and speed on time headway.}
	\label{fig:Delay_change}
\end{figure}

To investigate the influence of delay on the model's efficiency performance, we change the communication delay (no randomness) gradually from 0 to 1 s and output the corresponding stable (when both vehicles reach the same speed) time headway for a small-follow-small vehicle pair under different speeds in Fig. \ref{fig:Delay_change}. As we can see, if no delay exists (an unrealistic scenario), the time headway can be as low as 0.165 s when both vehicles travel with a speed of 120 km/h. Such an efficiency, however, can never be achieved in the real world due to unavoidable delays. When the communication delay is set to 0.1 s (a reasonable value in practice), the time headway of the proposed model can achieve 0.45 s, which is slightly smaller than the shortest headway (0.6 s) of CACC vehicles \cite{milanes2014modeling}, implying that our proposed model is efficient compared with commercially applied standards.

\begin{figure}[h!]
	\centering
	\subfloat[][PV: midsize, FV: small]{\resizebox{0.158\textwidth}{!}{
			\includegraphics[width=1\textwidth]{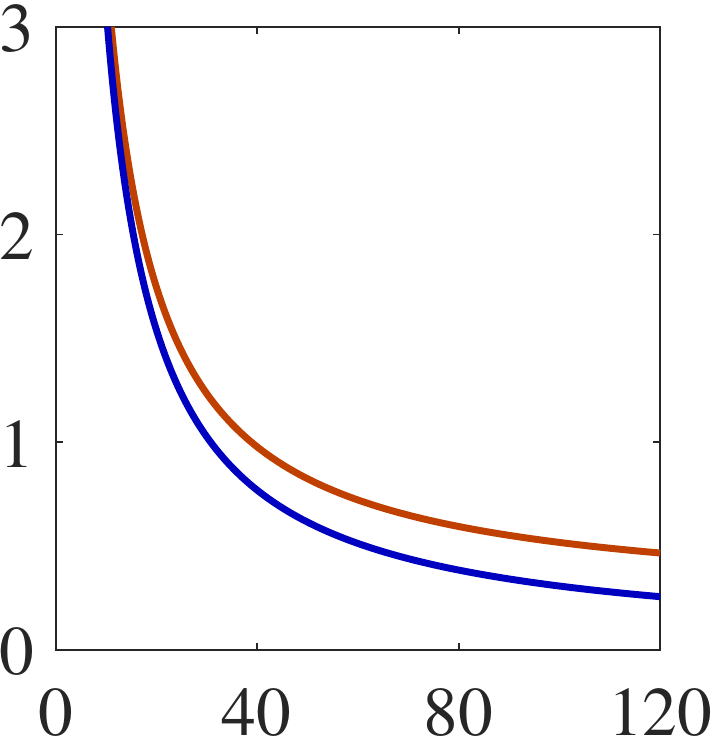}}
		\label{fig:C1}
	} 		
	\subfloat[][PV: large, FV: small]{\resizebox{0.158\textwidth}{!}{
			\includegraphics[width=1\textwidth]{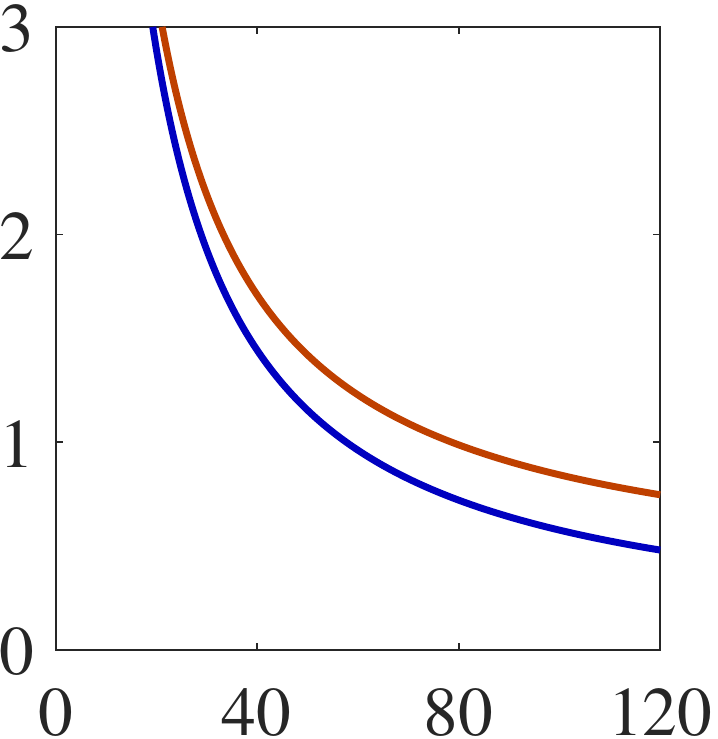}}
		\label{fig:C2}
	}
	\subfloat[][PV: large, FV: midsize]{\resizebox{0.158\textwidth}{!}{
			\includegraphics[width=1\textwidth]{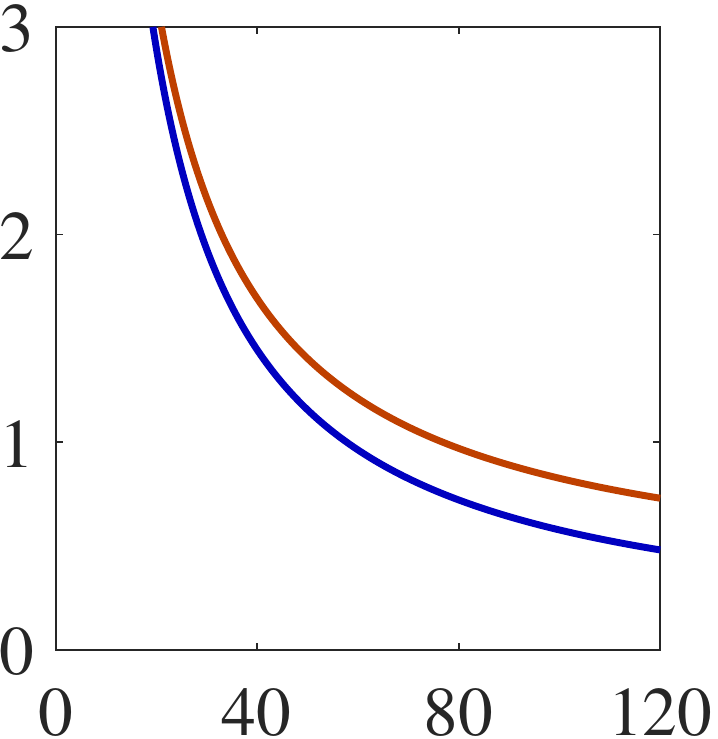}}
		\label{fig:C3}
	}\\[0.9ex] 
	
	\subfloat{\resizebox{0.185\textwidth}{!}{
			\includegraphics[width=1\textwidth]{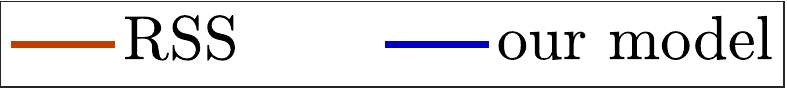}} 
		\label{fig:Cb}
	}\\[-1ex]
	\caption{Comparison between RSS and our model, x-axis: speed (km/h), y-axis: time headway (s).}
	\label{F:rss}
\end{figure}

To compare the efficiency of our model with the state-of-art research, we pick the representative RSS model  \cite{shalev2017formal} and compute the stable time headway of both models under different speeds. 
	RSS model 
	can guarantee absolute safety for disconnected AVs as our model does for CAVs. 
	As stated in Sec.\ref{S:litr}, the RSS model only considers a safety constraint at a full stop by assuming that FV always brakes more slowly than PV, hence it tends to keep a larger gap to guarantee safety when FV can brake faster than PV. Therefore, we choose small-follow-midsize, small-follow-large, and midsize-follow-large to see how much efficiency is sacrificed by RSS in order to achieve safety. In this experiment, we set the delay fixed at 0.1 s in both models, and the comparison results are shown in Fig. \ref{F:rss}. It is clear that our model with three complete safety constraints can achieve smaller headway especially when the speed is larger than 30 km/h. The average efficiency improvements for the three tested car-following pairs are about 17\%, 29\%, and 38\% respectively when the speed is 40, 80, and 120 km/h, respectively.

\subsection{Platoon with diversified car-following pairs}
\label{SS:platoon}
We design a platoon of 10 vehicles with three vehicle types to test the performance of the proposed car-following model. Table \ref{T:ord} shows the vehicle sequence and the corresponding vehicle type. The sequence is arranged so that the platoon includes all nine kinds of car-following pairs. All vehicles have a maximum speed of 22 m/s (about 80 km/h). Except for the first vehicle, all vehicles behave according to the car-following model. The first half trajectory of the leading vehicle is designed to show the scenario with fluctuating speeds, and the second half trajectory is designed to show the scenario of a hard brake. Note that no packet loss is considered so far.
\begin{table}[h!]
	\caption{Vehicle sequence and type}
	\begin{tabular}{cc|cc}
		\hline
		~~Sequence~~ & ~~Vehicle type~~~~~ & ~~~Sequence~~ & ~~Vehicle type~~ \\ \hline
		1     & small    & 6     & large    \\
		2     & small    & 7     & small    \\
		3     & midsize  & 8     & large    \\
		4     & midsize  & 9     & mid      \\
		5     & large    & 10    & small    \\ \hline
	\end{tabular} \vspace{-3mm}
	\label{T:ord}
\end{table}

\begin{figure}[h!]
	\centering
	\subfloat[][Trajectories]{\resizebox{0.45\textwidth}{!}{
			\includegraphics[width=1\textwidth]{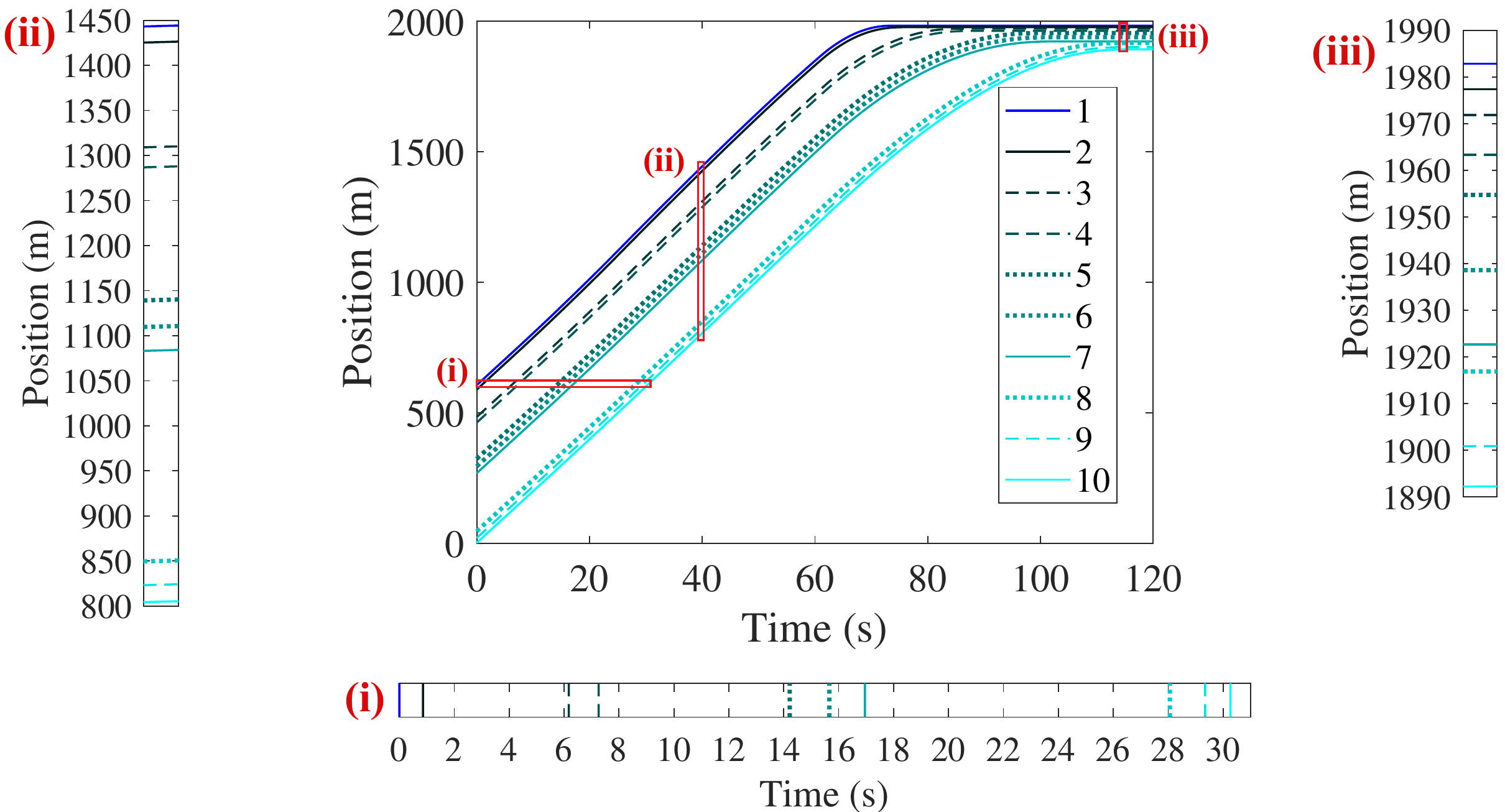}}
		\label{fig:1_tra}
	} 		
	
	\subfloat[][Speeds]{\resizebox{0.24\textwidth}{!}{
			\includegraphics[width=0.2\textwidth]{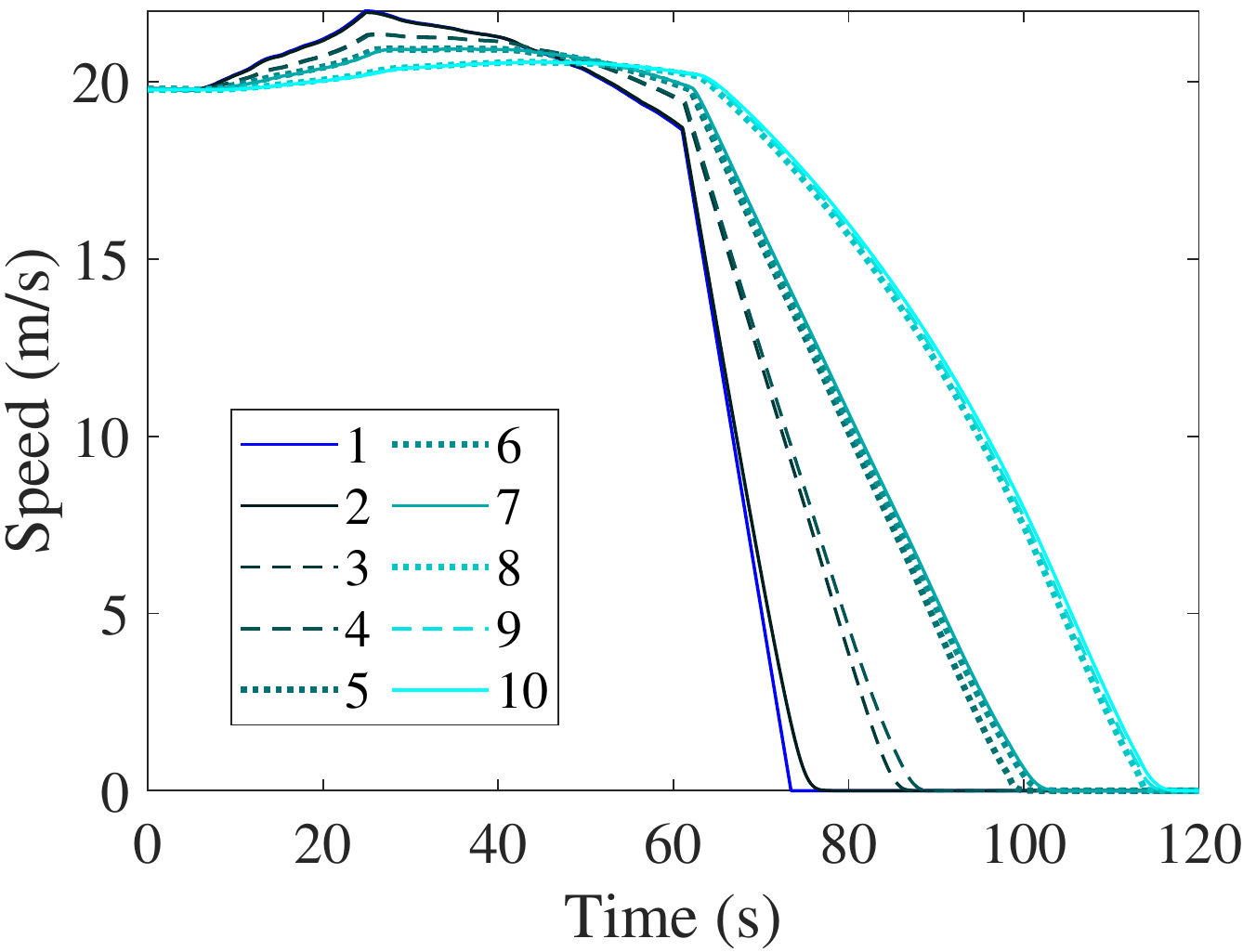}}
		\label{fig:1_speed}
	}
	\subfloat[][Accelerations]{\resizebox{0.24\textwidth}{!}{
			\includegraphics[width=1\textwidth]{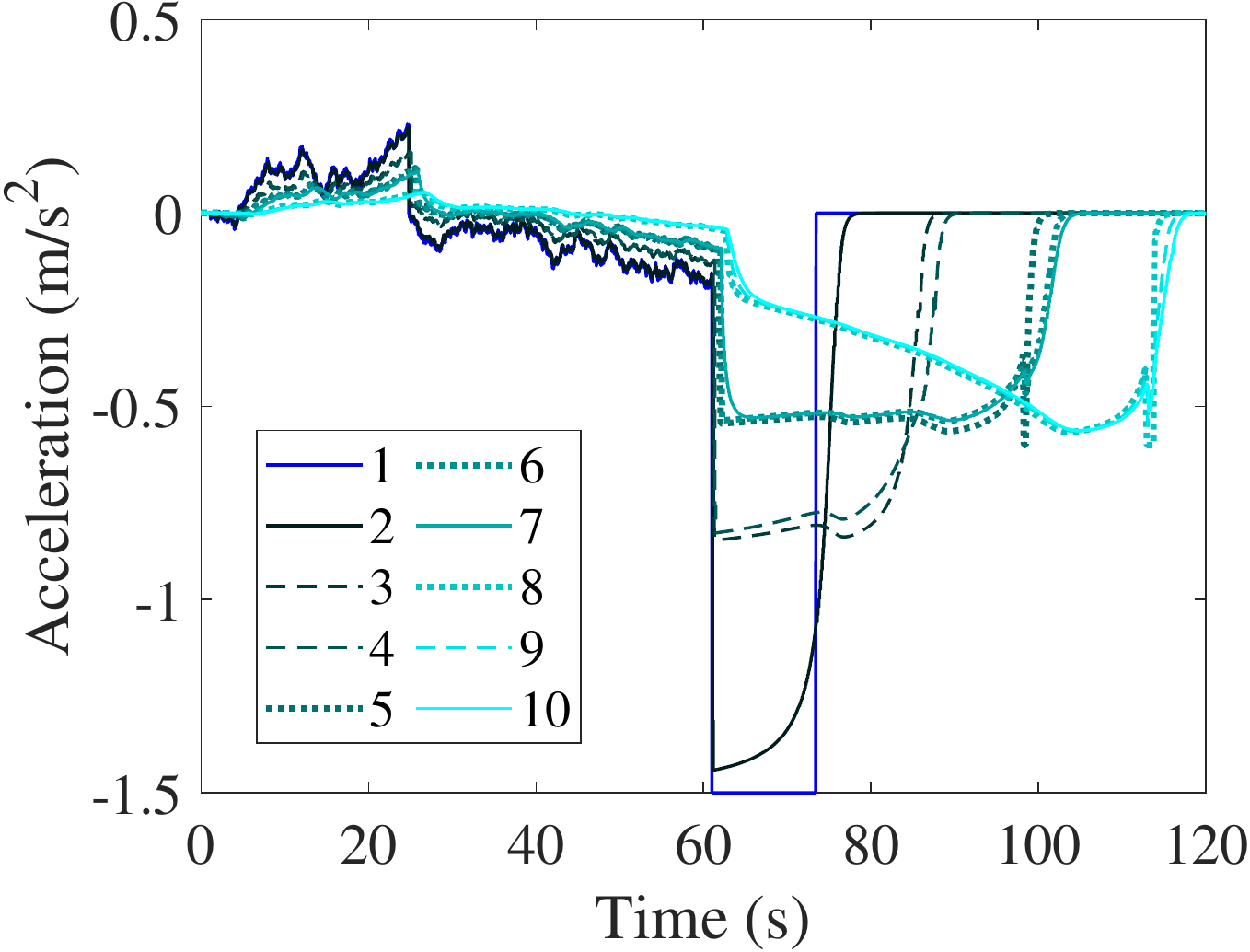}}
		\label{fig:1_acc}
	}
	\caption{A mixed vehicle platoon: small vehicle: ``---'', midsize vehicle: ``- - -'', and large vehicle: ``$\cdot\cdot\cdot$''.}
	\label{F:1_}
\end{figure}

Fig. \ref{F:1_}\subref{fig:1_tra}, \subref{fig:1_speed}, and \subref{fig:1_acc} show the result of trajectory, speed, and acceleration, respectively. Note that Fig. \ref{F:1_} \subref{fig:1_tra} also shows (i) time headway during driving, (ii) space headway during driving and (iii) space headway while stop.
We can see from Fig. \ref{F:1_} \subref{fig:1_tra} that safety is always guaranteed during the whole driving. When an FV has the same or a stronger braking ability compared to its PV, the space and time headway will be small during driving (time headway of around 1 s, indicating a high car-following efficiency), and their speeds will be similar according to Fig. \ref{F:1_} \subref{fig:1_speed}. On the contrary, the three large ``jumps" in space and time headway during driving happen when a midsize CAV follows a small CAV, a large CAV follows a midsize CAV, and a large CAV follows a small CAV, with a time headway of about 2.5, 3.5, and 5.5 s, respectively. In these three car-following pairs, the FV has a weaker braking ability and hence needs to keep a large headway from its PV considering its longer braking distance. Fig. \ref{F:1_}\subref{fig:1_acc} indicates that the acceleration fluctuation of the leading vehicle diminishes when propagating along the vehicle string, which implies the string stability of the proposed car-following model.

\subsection{Influence of packet loss}
In case of an unstable communication environment, CAVs may experience frequent packet losses. We assume that packet loss follows the Bernoulli distribution. That is, each information transmission event is independent. To \red{improve driving comfort and} reduce the acceleration fluctuations due to packet losses, we design four additional modules in the simulation, listed as follows.
\begin{enumerate}[label=\arabic*)]
	\item Once the needed information is missing, the CAV will use the neighboring information to conservatively estimate the lost information.
	\item If the needed missing information	is difficult to estimate, and the conservatively estimated bumper-to-bumper distance is enough, the CAV will continue to use the decision from its last decision moment.
	\item In case of a heavy packet loss rate ($>10\%$), CAVs will extend their $\kappa$ by 1 s to cope with the high uncertainty.
	\item In case of a heavy packet loss rate ($>10\%$), we also add an additional constraint $\ddot{x}_n^t \leq \ddot{x}_n^{t-\delta} + 0.1 \delta \overline{\ddot{x}}_n$, which limits the increment of acceleration during each $\delta$ within $0.1\delta\overline{\ddot{x}}_n$.
\end{enumerate}
These modules would significantly mitigate the fluctuation of acceleration and improve the comfort of passengers under the scenario with heavy packet loss, but at the sacrifice of efficiency to some degree.

\begin{figure}[h!]\vspace{-3mm}
	\centering
	\subfloat[][$1\%$ packet loss]{\resizebox{0.24\textwidth}{!}{
			\includegraphics[width=1\textwidth]{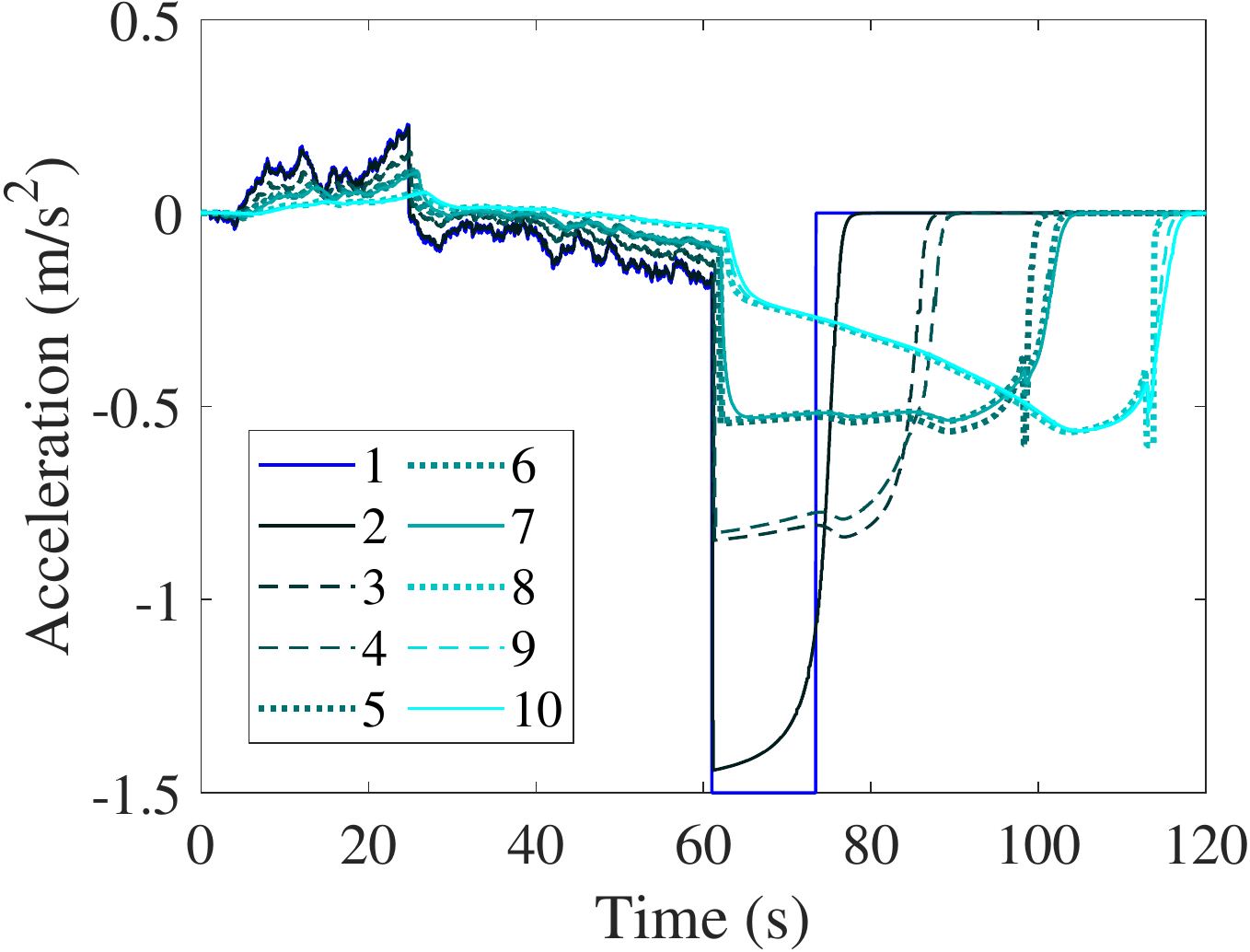}}
		\label{fig:0.01}
	} 		
	\subfloat[][$10\%$ packet loss]{\resizebox{0.24\textwidth}{!}{
			\includegraphics[width=0.2\textwidth]{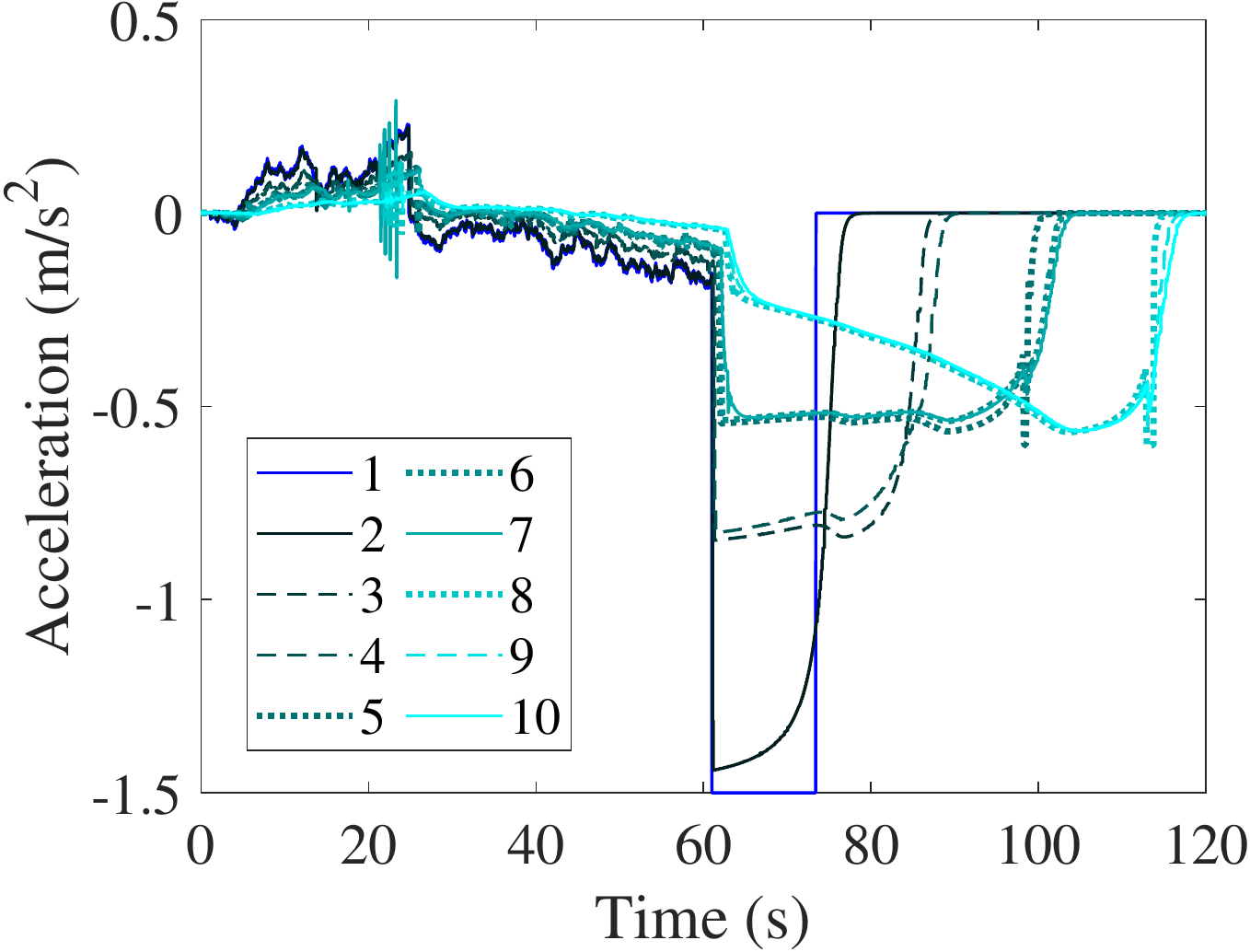}}
		\label{fig:0.1}
	}
	
	\subfloat[][$25\%$ packet loss]{\resizebox{0.2375\textwidth}{!}{
			\includegraphics[width=1\textwidth]{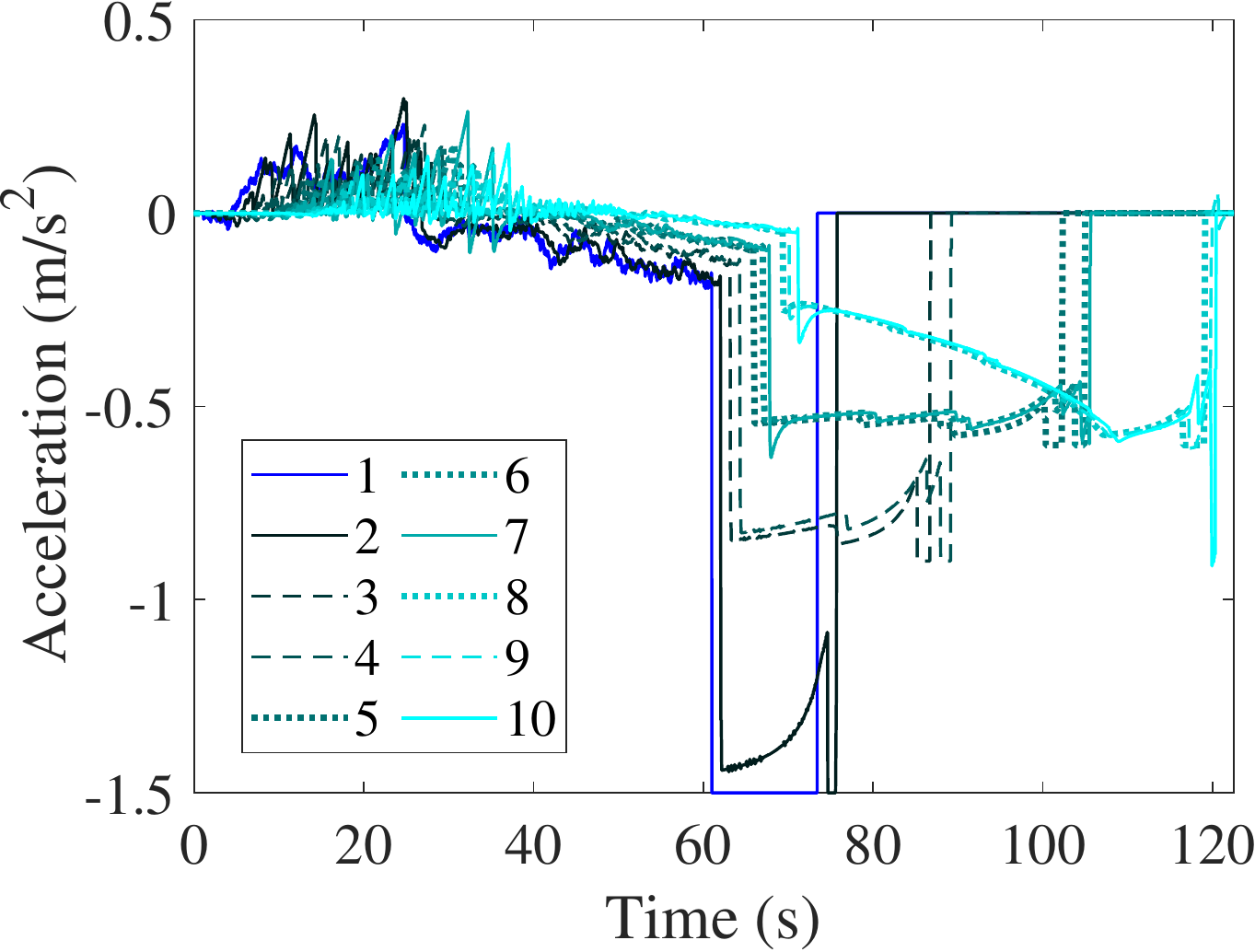}}
		\label{fig:0.25}
	} 		
	\subfloat[][$50\%$ packet loss]{\resizebox{0.24\textwidth}{!}{
			\includegraphics[width=0.2\textwidth]{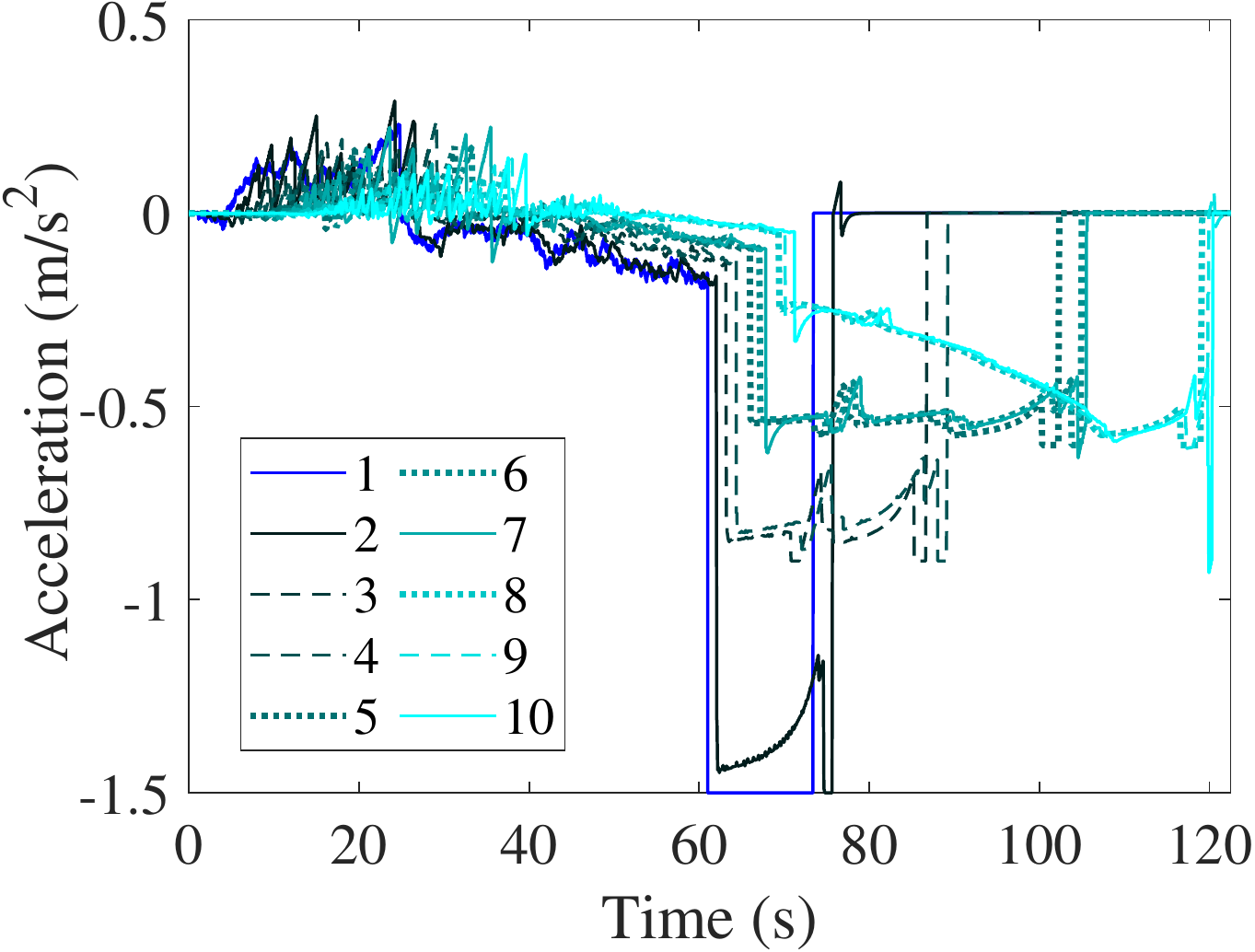}}
		\label{fig:0.5}
	}
	
	\caption{Influence of packet loss rate: small vehicle: ``---'', midsize vehicle: ``- - -'', and large vehicle: ``$\cdot\cdot\cdot$''.}
	\label{F:PL}
\end{figure}
Fig. \ref{F:PL} shows the results under different packet loss rates. The platoon configuration is the same with Sec. \ref{SS:platoon}. Note that Fig. \ref{F:PL} only demonstrates the acceleration and does not include the trajectory and speed because the main differences are reflected in the acceleration. No rear-end crashes happened in all scenarios. According to Fig. \ref{F:PL}, a low packet loss rate ($\leq 10\%$) brings little impact on vehicle accelerations. As the packet loss rate further increases, we see more fluctuations in the FV's acceleration, but such fluctuations are still under control and acceptable. 

\begin{figure}[h!]\vspace{-3mm}
	\centering
	\subfloat[][Trajectory]{\resizebox{0.24\textwidth}{!}{
			\includegraphics[width=1\textwidth]{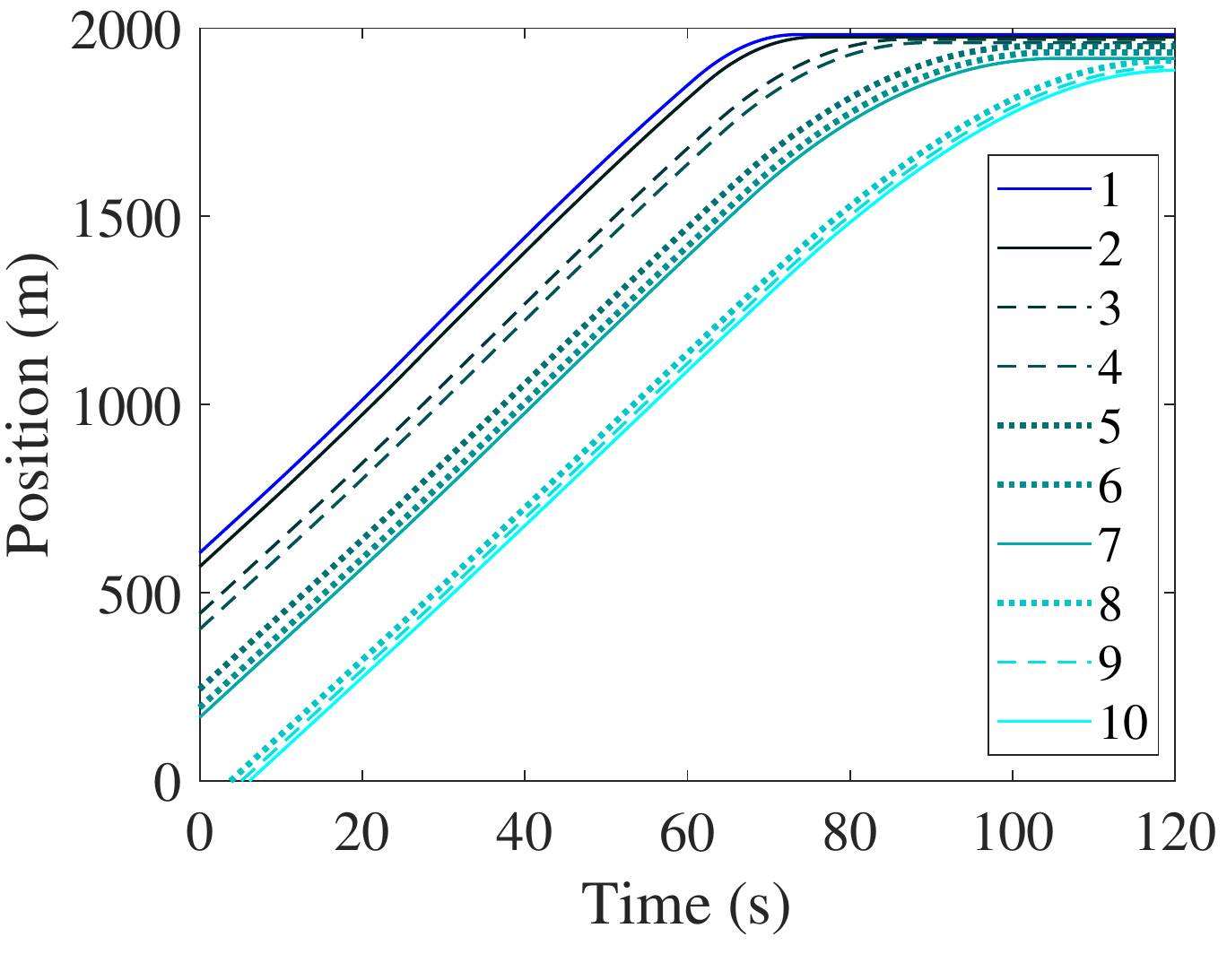}}
		\label{fig:tra_50}
	} 		
	\subfloat[][Max and min jerk]{\resizebox{0.24\textwidth}{!}{
			\includegraphics[width=1\textwidth]{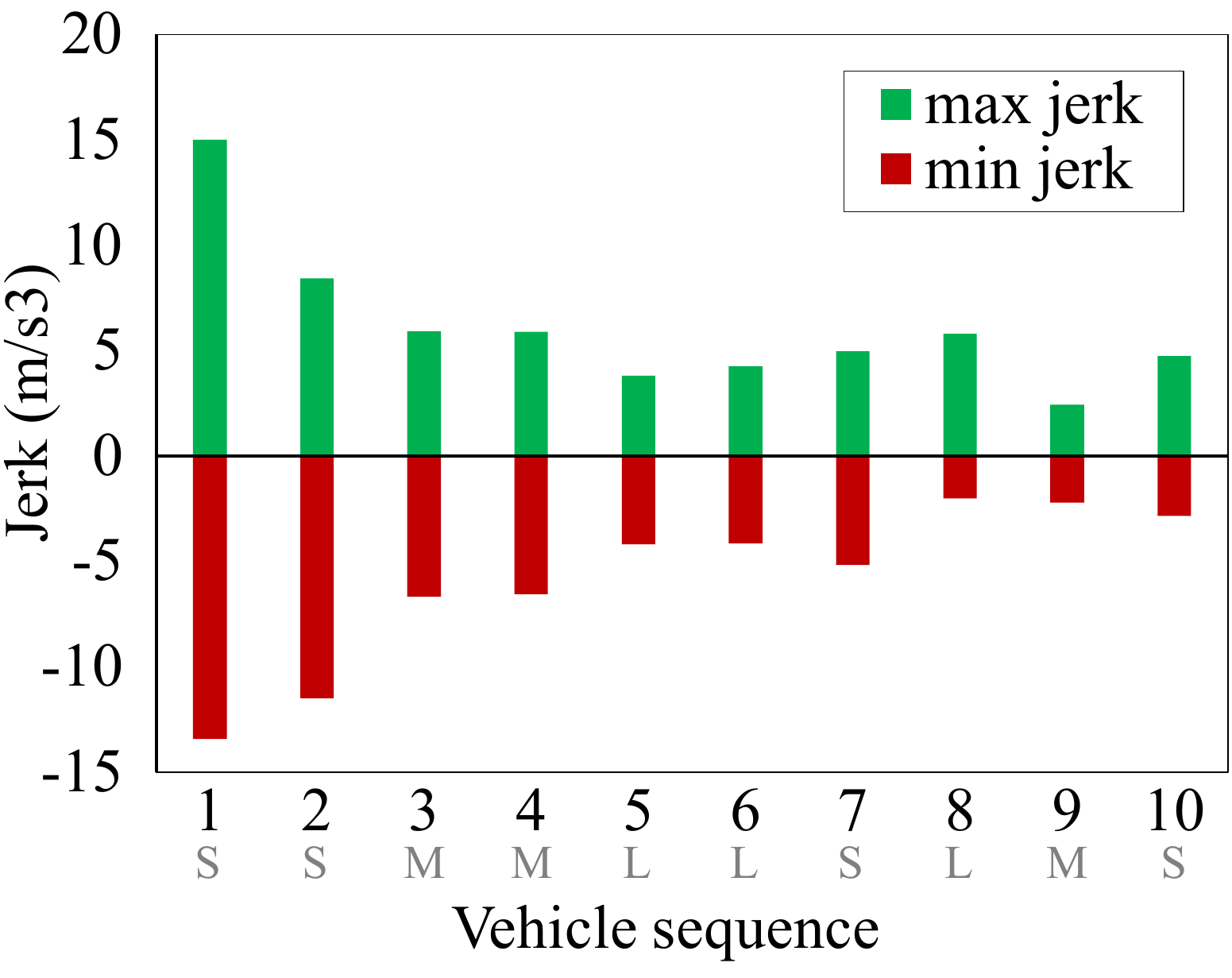}}
		\label{fig:Jerk_50}
	}
	\caption{Platoon performance with 50\% packet loss.}
	\label{F:per_50}
\end{figure}

We further show the details with 50\% packet loss in Fig. \ref{F:per_50}. Fig. \ref{F:per_50}\subref{fig:tra_50} is the vehicles' corresponding trajectories, and Fig. \ref{F:per_50}\subref{fig:Jerk_50} is their maximum and minimum jerk along the platoon (with 0.1s resolution).
As we can see, no crossing happens in vehicles' trajectories, which implies that the proposed car-following model is able to ensure safety during the whole driving process. Compared with Fig. \ref{F:1_}\subref{fig:1_tra} (trajectories with no packet loss), vehicles have to keep larges space due to the high packet loss. Jerk is the vehicle’s acceleration change rate with respect to time, and it is often used to measure driving comfort because it has a strong influence on the comfort of the passengers \cite{jacobson1980models}. One can see that even when the communication experience 50\% packet loss, the jerk still diminishes along the platoon, implying that the proposed model ensures driving comfort even when the communication environment is rather poor.

\section{CONCLUSION \red{AND OUTLOOK}}
With the popularity of Connected and Automated Vehicle (CAV) technology in recent years, the car-following models become possible to guide the CAVs to self-drive in the real world. As the most concerned issue for field application, safety has attracted much attention from car-following researchers. However, existing studies ignored the influence of discrete signal on the safety of CAV car-following. Safety-Oriented Car-Following (SOCF) models in the literature only consider the safety constraint in case of a hard brake really happens, and ignores the collision risk during the normal driving process. Such neglect could lead to a crash for CAVs with discrete signals.

In this paper, we propose a SOCF model that reproduces the discrete communication and decision making of CAVs. When formulating the model, we comprehensively consider the safety constraints during both normal driving process and a sudden braking process. We also incorporate mechanical delay into the communicated information to improve efficiency. Besides, we design four modules in the simulation to improve the driving comfort and platoon stability in case of high packet loss rates. Simulations show that each safety constraint plays an important role in guaranteeing the car-following safety. A simulation of platoon with diversified vehicle types shows the safety, efficiency, and string stability of the proposed model. Scenarios with different packet loss rates imply that the model can ensure driving safety and comfort in even poor communication environments.

\red{This paper only considers two vehicles (the preceding and the following) when formulating the car-following model. As stated above, this may enhance cyber security while sacrificing some efficiency. Evidence from the literature shows that collaborative driving can help improve sensing efficiency and resource utilization \cite{zhang2022collaborative}. An extension of the proposed SOCF model to collaborative driving, considering vehicle groups, is an exciting topic in the future. When scaling up the two-vehicle model to a platoon or a vehicle group model, the potential challenges that need to be addressed mainly include 1) how to make full use of additional information from extra vehicles and deal with the related communication and computing issues; 2) how to optimize the vehicle group's efficiency, comfort, and string stability with the safety guarantee; and 3) how to cope with the safety issues when the lane change is involved.}

\red{In addition, this paper assumes that each vehicle's communication and computing resource is sufficient. In case of poor communication and computing resources, the edge server may be necessary for improving the model performance \cite{yang2021edge, xiong2020intelligent, yang2020lessons, liu2019edge}. Future studies can investigate the combination of the proposed SOCF model and edge intelligence. Finally, although comfort is considered in the proposed model, it is accomplished through heuristic designs without theoretical guarantees. A recent study presented statistical models that describe automated vehicle driving comfort as a function of acceleration, jerk, and direction \cite{de2023standards}. Integrating the proposed SOCF model with such comfort models can theoretically guarantee driving comfort, which forms a promising research direction.}


%





\ifCLASSOPTIONcaptionsoff
  \newpage
\fi



\bibliographystyle{IEEEtran}
\bibliography{literature}
%



%

\begin{IEEEbiography}[{\includegraphics[width=1in,height=1.25in,clip,keepaspectratio]{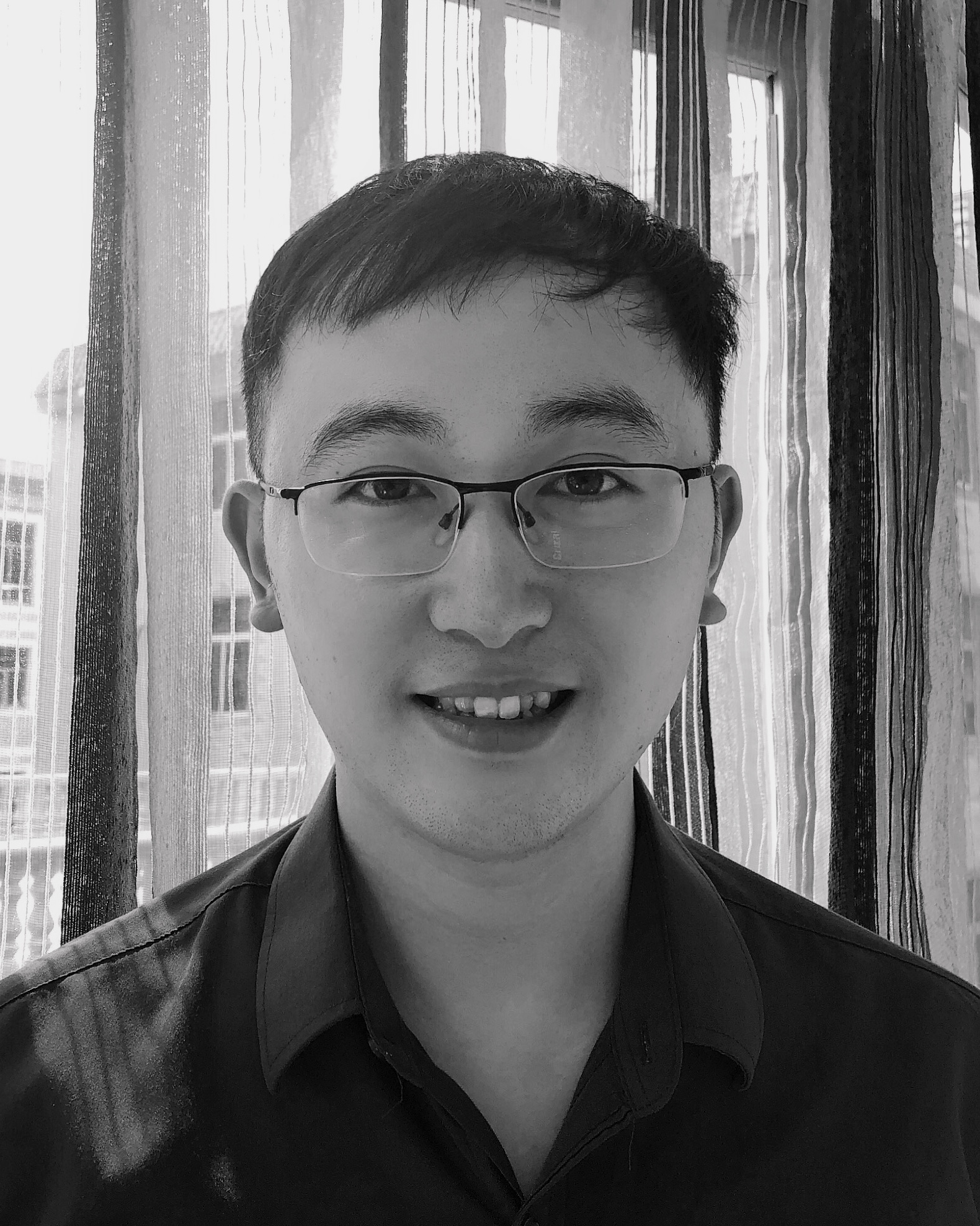}}]{DianChao Lin}
	was born in Fujian, China, in 1990. He received his B.Sc. and M.Sc. degrees in Traffic Engineering and Traffic Information Engineering \& Control from Tongji University, Shanghai, China, in 2013 and 2016, respectively, and the Ph.D. degree in Transportation Planning \& Engineering from New York University, NY, U.S.A., in 2021. He is currently an assistant professor with the School of Economics and Management, Fuzhou University, Fuzhou, China. His research interests include traffic management with economic schemes, connected vehicles, applications of game theory to automated vehicles, car-following of automated vehicles, non-motorized traffic behavior and multi-modal traffic flow.
\end{IEEEbiography}
\begin{IEEEbiography}[{\includegraphics[width=1in,height=1.25in,clip,keepaspectratio]{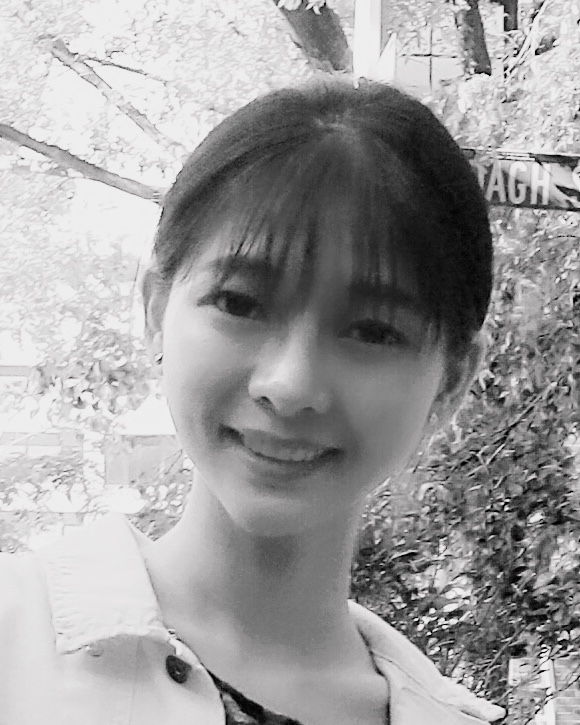}}]{Li Li} 
	was born in Jiangxi, China, in 1991. She received her B.Sc. and M.Sc. degrees in Traffic Engineering from Tongji University, Shanghai, China, in 2013 and 2016, respectively, and the Ph.D. degree in Transportation Planning \& Engineering from New York University, NY, U.S.A., in 2021. She is currently an assistant professor with the School of Civil Engineering, Fuzhou University, Fuzhou, China. Her research interests include network traffic management, connected vehicles, automated vehicles, shared mobility, traffic signal control, car following, lane change, and economic strategies in traffic management. ~~~~~~~~
\end{IEEEbiography}





\end{document}